%% file: main.tex
\documentclass[%
 reprint,
groupedaddress,
 amsmath,amssymb,
 aps,
]{revtex4-1}
\usepackage{amssymb}
\usepackage{amsthm}
\usepackage{amsmath}
\usepackage{etoolbox}
\usepackage{bbold}
\usepackage{ulem,url}
\usepackage{graphicx}
\usepackage{dcolumn}
\usepackage{bm}
\usepackage{hyperref}
\usepackage[mathlines]{lineno}
\usepackage{cleveref}
\usepackage{color}

\newcommand{\KL}[2]{\ensuremath{\mathrm{D}(#1|#2)}}
\newcommand{\KLtext}{Kullback-Leibler}
\newcommand{\KLmetric}{KL metric}
\newcommand{\MMD}{\ensuremath{\mathcal{L}_{MMD}}}

\newcommand{\TOKYO}{\texttt{ibmq\_20\_tokyo}}
\newcommand{\BOBO}{\texttt{ibmq\_boeblingen}}

\newcommand{\VAL}{\texttt{ibmq\_valencia}}
\begin{document}

\preprint{APS/123-QED}

\title{Error-mitigated data-driven circuit learning on noisy quantum hardware}
\thanks{This manuscript has been authored by UT-Battelle, LLC, under Contract No. DE-AC0500OR22725 with the U.S. Department of Energy. The United States Government retains and the publisher, by accepting the article for publication, acknowledges that the United States Government retains a non-exclusive, paid-up, irrevocable, world-wide license to publish or reproduce the published form of this manuscript, or allow others to do so, for the United States Government purposes. The Department of Energy will provide public access to these results of federally sponsored research in accordance with the DOE Public Access Plan.}%

\author{Kathleen E. Hamilton}
\author{Raphael C. Pooser}
\affiliation{Computer Science and Engineering Division, One Bethel Valley Road, Oak Ridge, TN USA}

\date{\today}

\begin{abstract}
Application-inspired benchmarks measure how well a quantum device performs meaningful calculations.  In the case of parameterized circuit training, the computational task is the preparation of a target quantum state via optimization over a loss landscape.  This is complicated by various sources of noise, fixed hardware connectivity, and for generative modeling, the choice of target  distribution.  Gradient-based training has become a useful benchmarking task for noisy intermediate scale quantum computers because of the additional requirement that the optimization step uses the quantum device to estimate the loss function gradient.  In this work we use gradient-based data-driven circuit learning to benchmark the performance of several superconducting platform devices and present results that show how error mitigation can improve the training of quantum circuit Born machines with $28$ tunable parameters.
\begin{description}
\item[Usage]
Secondary publications and information retrieval purposes.
\item[PACS numbers]
May be entered using the \verb+\pacs{#1}+ command.
\end{description}
\end{abstract}

\pacs{Valid PACS appear here}
\keywords{quantum machine learning, error mitigation}
\maketitle



\section{Introduction}
\label{sec:introduction}
\input{introduction.tex}

\section{Methods}
\label{sec:methods}
\input{methods.tex}

\section{Error mitigation in metric evaluation}
\label{sec:experiment0}
\input{experiment0.tex}

\section{Error mitigation in gradient training}
\label{sec:experiment1}
\input{experiment1.tex}


\section{Discussion}
\label{sec:discussion}
\input{discussion.tex}

\section{Conclusions}
\label{sec:conclusions}
\input{conclusions.tex}

\section{Acknowledgements}
The authors would like to thank Ryan Bennink for helpful discussions about matrix-based error mitigation methods, Eugene Dumitrescu for general comments about error mitigation,  Vicente Leyton-Ortega for general discussions about DDCL and Nathan Wiebe for informative discussions about noisy channel coding.

This work was supported as part of the ASCR Quantum Testbed Pathfinder Program at Oak Ridge National Laboratory under FWP \#ERKJ332.  This research used quantum computing system resources of the Oak Ridge Leadership Computing Facility, which is a DOE Office of Science User Facility supported under Contract DE-AC05-00OR22725. Oak Ridge National Laboratory manages access to the IBM Q System as part of the IBM Q Network.

\bibliographystyle{unsrt}
\bibliography{qcbm_II_lit.bib}
\appendix
\input{appendix.tex}
\end{document}

%% file: introduction.tex
The field of quantum machine learning covers a diverse range of topics, from utilizing quantum computing to speed up classical model training  to using quantum circuits as analogues of classical models.  Parameterized quantum circuits have become a popular approach to construct general trainable quantum models. These circuits can find use in state preparation for algorithms compatible with noisy intermediate scale quantum (NISQ) hardware.  The training training process for these circuits is similar to the variational quantum eigensolver, and it can be supervised, which has been used for classifiers \cite{mitarai2018quantum,havlivcek2019supervised}, unsupervised, and it can be used for generative models \cite{benedetti2019generative,liu2018differentiable}. 


In the absence of noise, circuits with few qubits and entangling layers are able to model general discrete and continuous distributions.  
However, realizing these modeling properties on current NISQ hardware faces a number of challenges.  We previously presented results on generative learning as a benchmark for circuits deployed on a superconducting qubit system developed by IBM \cite{hamilton2019generative}.  
In the present work we further explore how the classical optimizer Adam \cite{kingma2014adam} in concert with error mitigation can compensate for various sources of noise inherent in NISQ devices during circuit training, such as readout error and systematic error in two qubit gates. We show two approaches that apply error mitigation for state preparation and measurement (SPAM) errors to quantum circuit Born machines (QCBMs). In the first case, we use error mitigation in post-processing to estimate the learned distribution, and use the mitigated values to calculate error-mitigated performance metrics such as the \KLtext{}(KL) divergence. In the second case, the error mitigation scheme runs alongside the optimizer at each training step, post-processing the results for the loss function gradient before passing it on to the optimizer for evaluation at the next training step. We tested two error mitigation schemes, one that characterizes and corrects for general readout error, and another method which characterizes the errors associated with the specific QCBM circuit under study. Both methods rely on constructing a correlation matrix of assignment errors, using this matrix as a characterization of the noise in a given device at one point in time, and relying on this characterization throughout the training of a circuit.

Combining the correlation matrix inversion with training, and with distribution and metric evaluation resulted in improved performance for three different output distributions on multiple hardware platforms. We constructed QCBM circuits to learn the Bars and Stripes distribution encoded on four qubits~\cite{benedetti2019generative,liu2018differentiable,hamilton2019generative} and two Poisson distributions to demonstrate the capabilities and limitations of the error-mitigated learning scheme. Using the Qiskit programming library \cite{Qiskit} we tested the benchmark on several quantum processor units (QPUs): the IBM Tokyo (\TOKYO), Boeblingen (\BOBO), and Valencia (\VAL) processors.  The connectivity and error rates of each chip resulted in different levels of in metric performance, allowing us to use the benchmark as a point of comparison. In addition to metric performance, the degree to which error mitigation improves the metric can be used as a verification of the degree to which the errors present on the device are systematic and correctable. In comparison with other error mitigation schemes, such as Richardson extrapolation~\cite{temme2017error,li2017efficient,havlivcek2019supervised} or probabilistic cancellation~\cite{temme2017error}, the matrix inversion method seeks to mitigate primarily readout errors, but it treats the entire circuit as a black box within which errors can occur at any point.

%% file: methods.tex
\begin{figure}[htbp]
  \centering
  \includegraphics[width=\columnwidth]{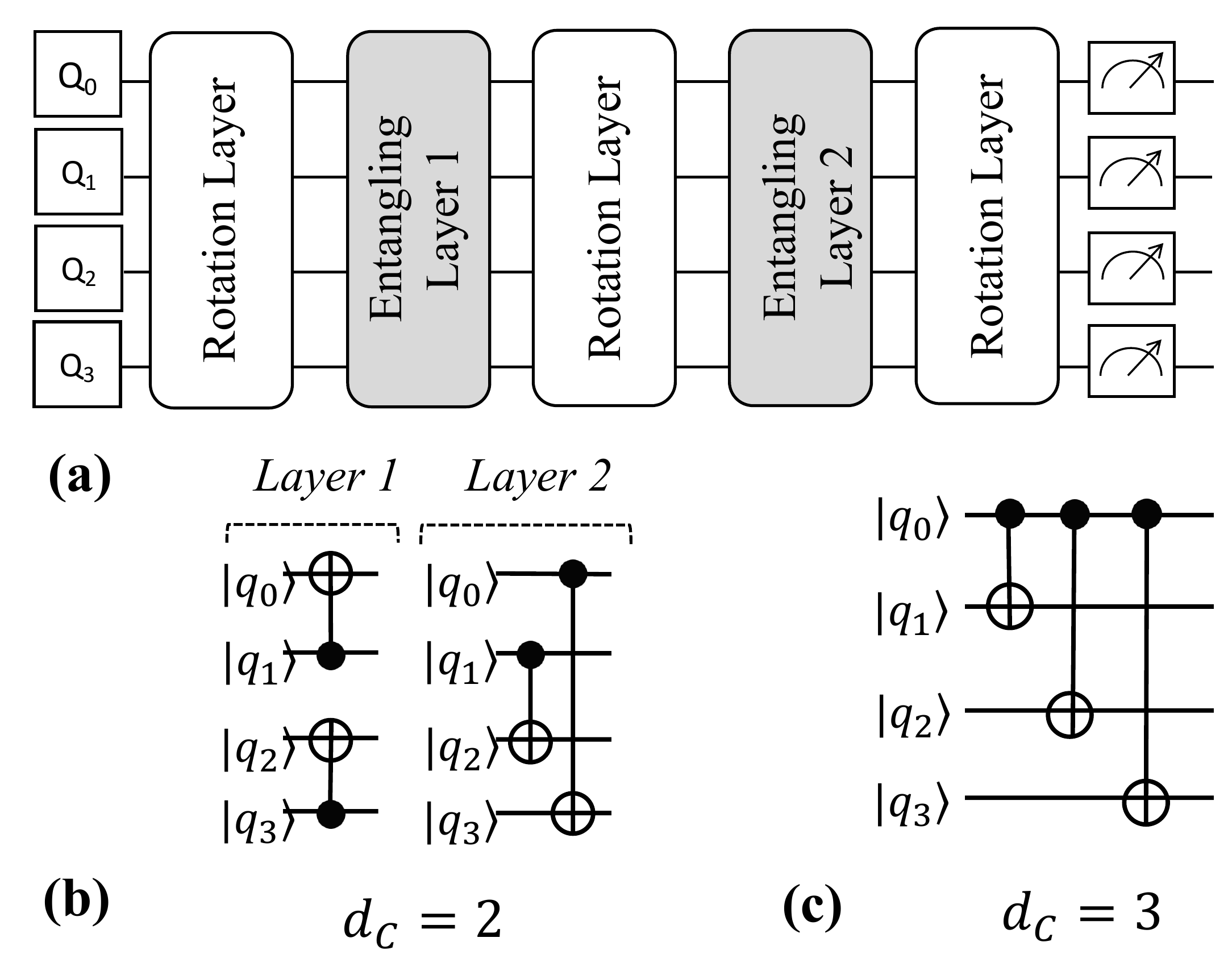}
  \caption{(a) The basic bi-layer circuit ansatz for DDCL \cite{kandala2017hardware,benedetti2019generative,liu2018differentiable,hamilton2019generative,leyton2019robust}, (b) $d_C = 2$ entangling layer and (c) $d_C=3$ entangling layer used previously in \cite{hamilton2019generative}.}
  \label{fig:ansatz_entanglers}
\end{figure}

Here we report on the training of several QCBM models using the general bi-layer parameterized circuit ansatz \cite{kandala2017hardware,liu2018differentiable,benedetti2019generative,hamilton2019generative} constructed with alternating layers of parameterized rotation gates ($R_X$ and $R_Z$) and layers of two-qubit entangling gates (CNOTs).  All circuits in this paper have $28$ trainable parameters (single-qubit gates), but the number of CNOTs in each circuit is dependent on the entangling layer design as shown in \Cref{fig:ansatz_entanglers}.  Circuits with 2 CNOTs per layer are denoted $d_C=2$, and circuits with 3 CNOTs per layer are denoted $d_C=3$. 

\subsection{Error mitigation method}
For a $N$-qubit circuit, we construct \textit{assignment error matrices} (AEM): $2^N \times 2^N$ real-valued, non-negative matrices ($\mathbb{K}_{kernel}$) that are assumed to be invertible and $kernel$ indicates one of $3$ classes of AEM.  The first class of AEM ($\mathbb{K}_{\mathbb{1}}$) are trivial; they account for no noise, require no circuits be evaluated on the hardware and are simply the $2^N \times 2^N$ identity matrix $\mathbb{1}$.  

\begin{figure}[htbp]
  \centering
  \includegraphics[width=\columnwidth]{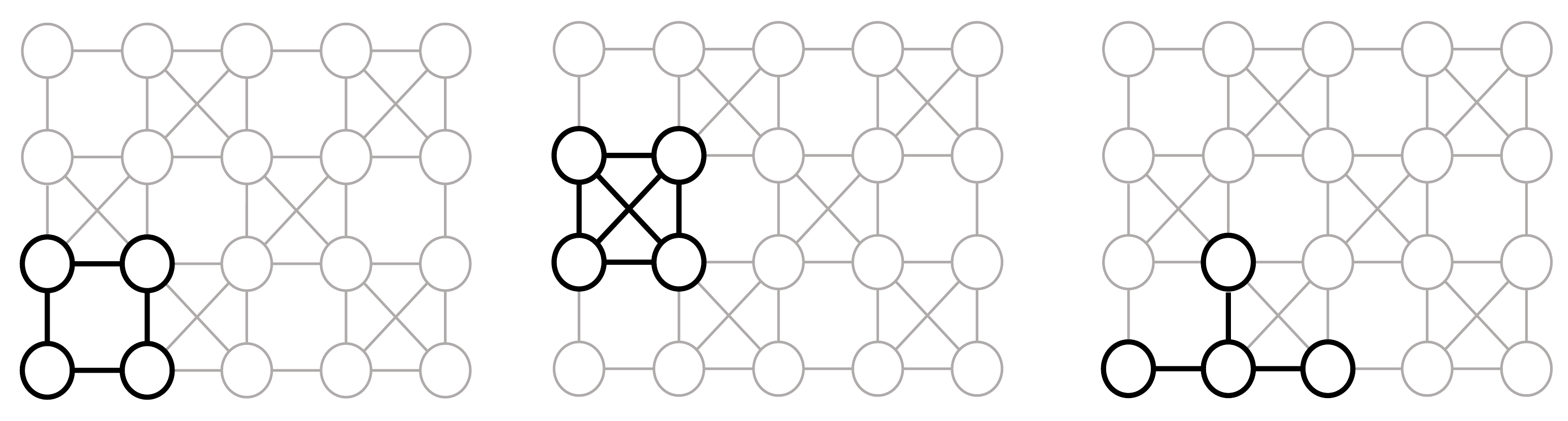}
  \caption{Qubits used on \TOKYO:  (Left) $P_A$ (Center) $P_B$ (Right) $T_1$.}
  \label{fig:tokyo_hardware_qubits}
\end{figure}

\begin{figure}[htbp]
  \centering
  \includegraphics[width=\columnwidth]{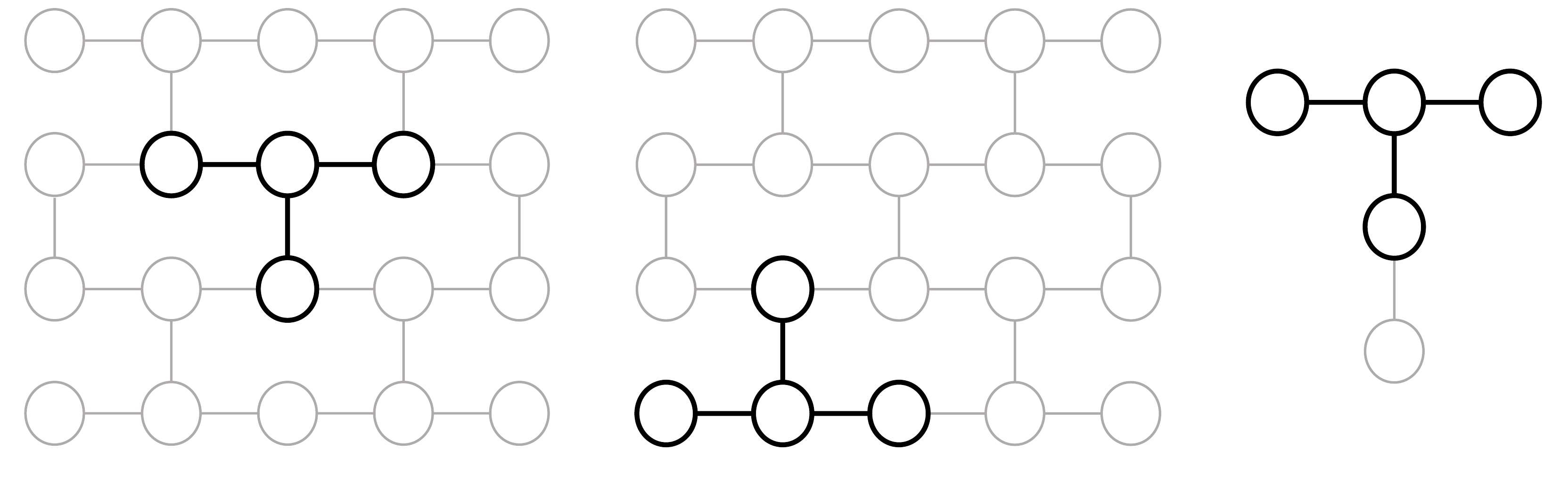}
  \caption{Qubits used on: \BOBO{} and \VAL. (Left) $T_0$ on \BOBO, (Center) $T_1$ on \BOBO, (Right) \VAL{} has only one unique tree layout where the root has degree 3.}
  \label{fig:other_hardware}
\end{figure}

The remaining $2$ classes of AEM, $hw$ and $circ$ require hardware execution of $2^N$ quantum circuits.  The entries $\mathbb{K}_{kernel}(i,j)$ denote the probability that a QPU preparing the target state $i$ will measure state $j$.  The diagonal entries $\mathbb{K}_{kernel}(i,i)$ are the probabilities that the target state is returned and off-diagonal elements $\mathbb{K}_{kernel}(i,j)$ are the probabilities that another state is returned. In subsequent sections of the paper, the arguments $i,j$ are omitted from labels when it is clear the entire matrix is used as a label or otherwise all elements are to be considered.  Once executed the circuits are sampled at a shot size of $n_s=4096$ which is sufficient to resolve readout errors and reduces statistical error.  In \cite{kandala2017hardware} (Supp. Material) it is noted that assignment error correction assumes independent error, but correlated errors are possible (e.g. cross-resonance errors), thus any $\mathbb{K}$ is executed on localized subsets of qubits.  For the remainder of this paper we will restrict our choice of qubits to the specific subsets shown in \Cref{fig:tokyo_hardware_qubits,fig:other_hardware} which are used to construct $\mathbb{K}_{kernel}$ and also to train the individual QCBM.  

AEM that mitigate readout gate errors are denoted ($\mathbb{K}_{hw}$).  The $2^4=16$ shallow circuits used to construct $\mathbb{K}_{hw}$ contain only un-parameterized $X$-rotation gates.  The third class of AEM ($\mathbb{K}_{circ}$) are generated using the specific QCBM circuits (see \Cref{fig:ansatz_entanglers}(a)) with the goal of mitigating not just readout gate errors, but also general errors from two-qubit gates and the parameterized single qubit gates.  The rotational parameters needed for full reconstruction of the AEM are determined analytically and are dependent on the entangling layer gates.  Further detail about the AEMs used in the main text are given in \Cref{appendix:AEM}.

When the final state prepared from a circuit is sampled, the counts returned in a state ($c(|x_i\rangle)$) are converted into probabilities using the shot size ($q_i = c(|x_i \rangle)/n_s)$. Through error mitigation we remove erroneous counts in three steps. The vector of counts $c(\mathbf{x}))$ is multiplied by an inverted AEM: ($c^{\prime}(\mathbf{x}) =\mathbb{K}^{-1} c(\mathbf{x}))$.  Any negative counts are set to $0$.  The remaining counts are used to define an effective shot size $n_s^{\prime}$ which normalizes the \textit{mitigated distribution} $q^{\prime}(\mathbf{x}) = c^{\prime}(\mathbf{x})/n_s^{\prime}$.

If a circuit is not optimized for hardware deployment then compilation for hardware will add additional CNOT gates.  In the main text we consider only circuits that can be compiled with the minimum number of CNOTS ($2d_C$). We consider the case of training a non-optimal circuit on hardware in \Cref{appendix:SWAP_noise}.

In \Cref{sec:experiment0} we only incorporate error mitigation into the \KLmetric{} evaluation.  In \Cref{sec:experiment1} error mitigation is incorporated into DDCL using an AEM and the resulting mitigated distributions are used to evaluate the loss and gradient of the loss in training.  The AEM is generated once at the start of training, stored, and used throughout a DDCL workflow. Any time a state prepared by a circuit is sampled we define a mitigated distribution using the stored AEM.

\subsection{Data-driven Circuit Learning}
Data driven circuit learning (DDCL) for generative modeling trains a parameterized circuit ansatz to prepare a target distribution.  There are many different loss functions that can be utilized in training, and the optimization can be gradient-based or gradient-free.  Recent studies have used: clipped log-likelihoods \cite{benedetti2019generative}, the Jensen-Shannon divergence \cite{leyton2019robust}, the Stein divergence \cite{coyle2019born}.  Following the work of Ref \cite{liu2018differentiable} and our earlier studies \cite{hamilton2019generative} we use the maximum mean discrepancy (MMD) loss function with radial basis function kernels ($\sigma=0.1$) and gradient-based optimization with Adam \cite{kingma2014adam} using code adapted from \cite{Liu_github}.  The circuit gradient is evaluated using the parameter shift rule \cite{schuld2019evaluating,li2017efficient,liu2018differentiable}.  The parameter shift rule defines an exact expression for the circuit gradient, but device noise and sampling noise reduce any gradient evaluation on hardware to an estimation. Now with the addition of error mitigation this noisy estimate can be distorted due to the amplitude of a state in the mitigated distribution becoming zero. 

\begin{figure}[htbp]
  \centering
  \includegraphics[width=\columnwidth]{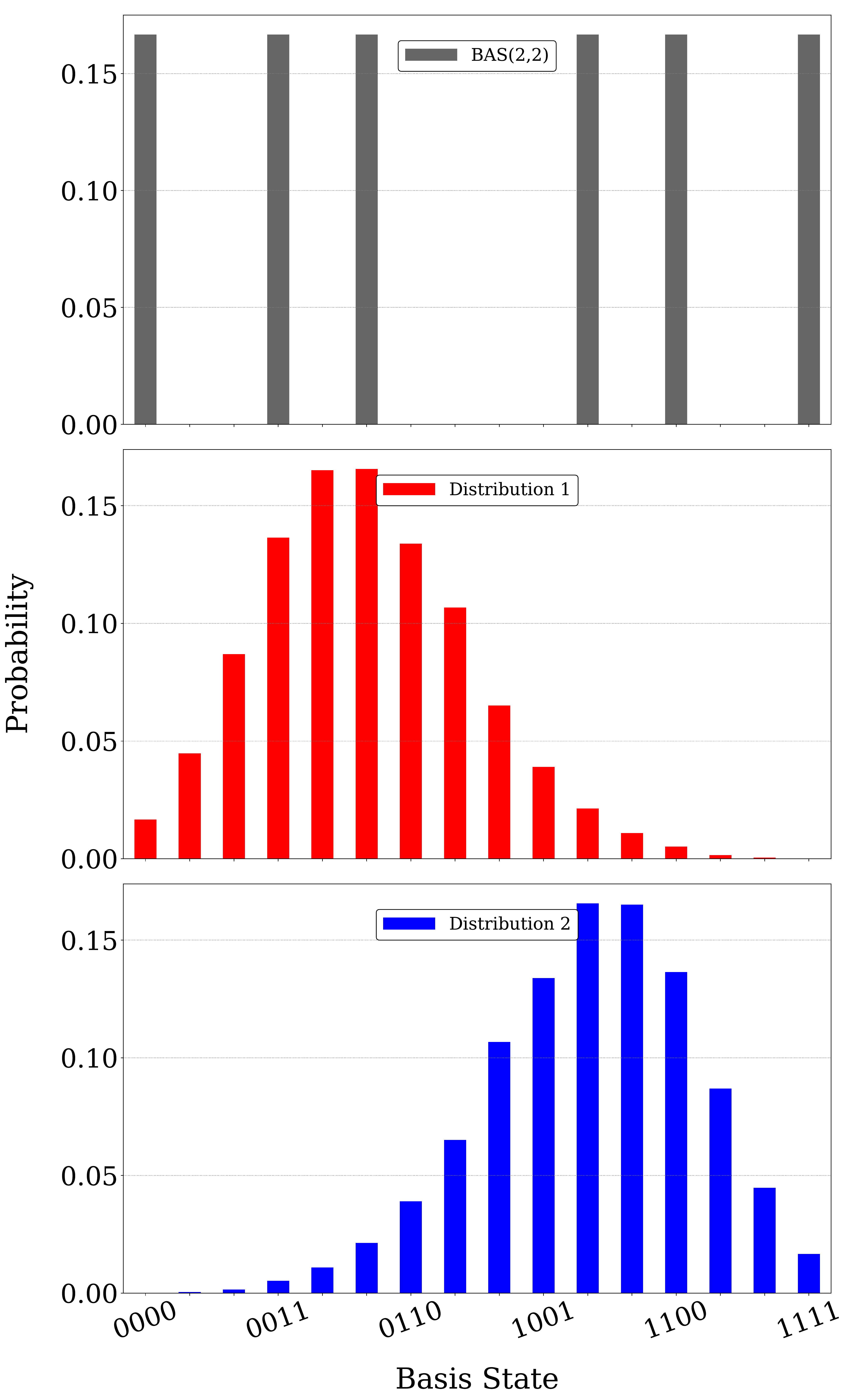}
  \caption{$3$ target distributions $p(\mathbf{x})$ mapped onto the $2^4=16$ basis states of a $4$-qubit circuit:  (Top) discrete Bars and Stripes distribution (Middle) Distribution $1$ is a discretized Poisson distribution, (Right) Distribution $2$ is constructed from Distribution $1$ found by applying \texttt{numpy.fliplr()}.}
  \label{fig:BAS}
\end{figure}

All target distributions are mapped onto the $16$ computational basis states of a $4$ qubit circuit.  In \Cref{sec:experiment0,sec:experiment1} the target distribution is the Bars and Stripes benchmark distribution (\Cref{fig:BAS}) (top panel), which was recently shown to be difficult to prepare on a quantum device due to its high entanglement entropy \cite{benedetti2019generative}.  In \Cref{sec:discussion} we expand our study to a set of non-uniform distributions (\Cref{fig:BAS} middle, bottom panels).

There are many loss functions used in training generative models \cite{mohamed2016learning,theis2015note} and a growing number of performance metrics  \cite{benedetti2019generative,hamilton2019generative} to evaluate the performance of DDCL.  We use the \KLtext{}(KL) divergence,
\begin{equation}
    \KL{p}{q^{\prime}}=\sum_i p(x_i) \log{\left(\frac{p(x_i)}{q^{\prime}(x_i)}\right)},
\end{equation}
to quantify the distance between a target distribution $p(\mathbf{x})$ and a measured one $q^{\prime}(\mathbf{x})$ and thus as a metric for performance of a QCBM.  The distributions $q^{\prime}(\mathbf{x})$ may be mitigated by non-trivial AEM ($\mathbb{K}_{hw,circ}$) or by the identity matrix ($\mathbb{K}_{\mathbb{1}}$).  

The \KLmetric{} diverges if $q^{\prime}(x_i)=0$ where $p(x_i)\neq 0$, but it is commonly used in the literature and allows us to compare our QCBM performance to other contemporary studies  \cite{benedetti2019generative,leyton2019robust,liu2018differentiable}.  To avoid the shortcoming of the \KLmetric{}, in training we use \MMD{} as a cost function. When $q^{\prime}(\mathbf{x})=p(\mathbf{x})$ both \KLmetric{} and \MMD{} vanish and multiple studies~\cite{benedetti2019generative,hamilton2019generative} show that optimizing a separate loss function yields good performance in minimizing the \KLmetric{} divergence.
This allows us the freedom to study how incorporating error mitigation in different parts of the DDCL workflow affect the overall QCBM performance. 

%% file: experiment0.tex
Here we present an initial study of how the assignment error matrix can be used as a post-processing technique to improve a given performance metric.  The ${d_C=2,3}$ circuit classes were used to construct several QCBMs that were trained on the $20$-qubit Tokyo superconductiong qubit system (\TOKYO) via gradient-based DDCL training \cite{liu2018differentiable,hamilton2019generative} for a fixed number ($25$) of Adam steps. The circuit parameters $\{\mathbf{\theta}\}$ are initialized with the same set of random starting values $\{\theta_i\}$ and during training the shot size is fixed at $2048$.  The AEM were generated May 8--17 2019, the trained circuit parameters $\{\theta_f\}$ were obtained on \TOKYO{} May 19--25 2019, and the un-mitigated distributions were generated May 21--25 2019.

The circuit parameters recorded during training are used to generate multiple distributions at each step.  For each set of parameters, a batch of $5$ identical circuits were sent to the hardware and evaluated with $2048$ shots.  These counts were used to construct a composite distribution with an effective shot size of $10240$.  

To study the effect of hardware AEM ($\mathbb{K}_{hw}$), a mitigated distribution was generated from the composite distribution.  From this mitigated composite distribution we then draw a  set of sub-samples to re-evaluate and provide an error-mitigated \KLmetric{}.  This procedure was repeated with the circuit AEM ($\mathbb{K}_{circ}$).  The training, distribution evaluation, and generation of each $\mathbb{K}$ are all done using the same hardware qubits.

\begin{figure}[htbp]
  \includegraphics[width=\columnwidth]{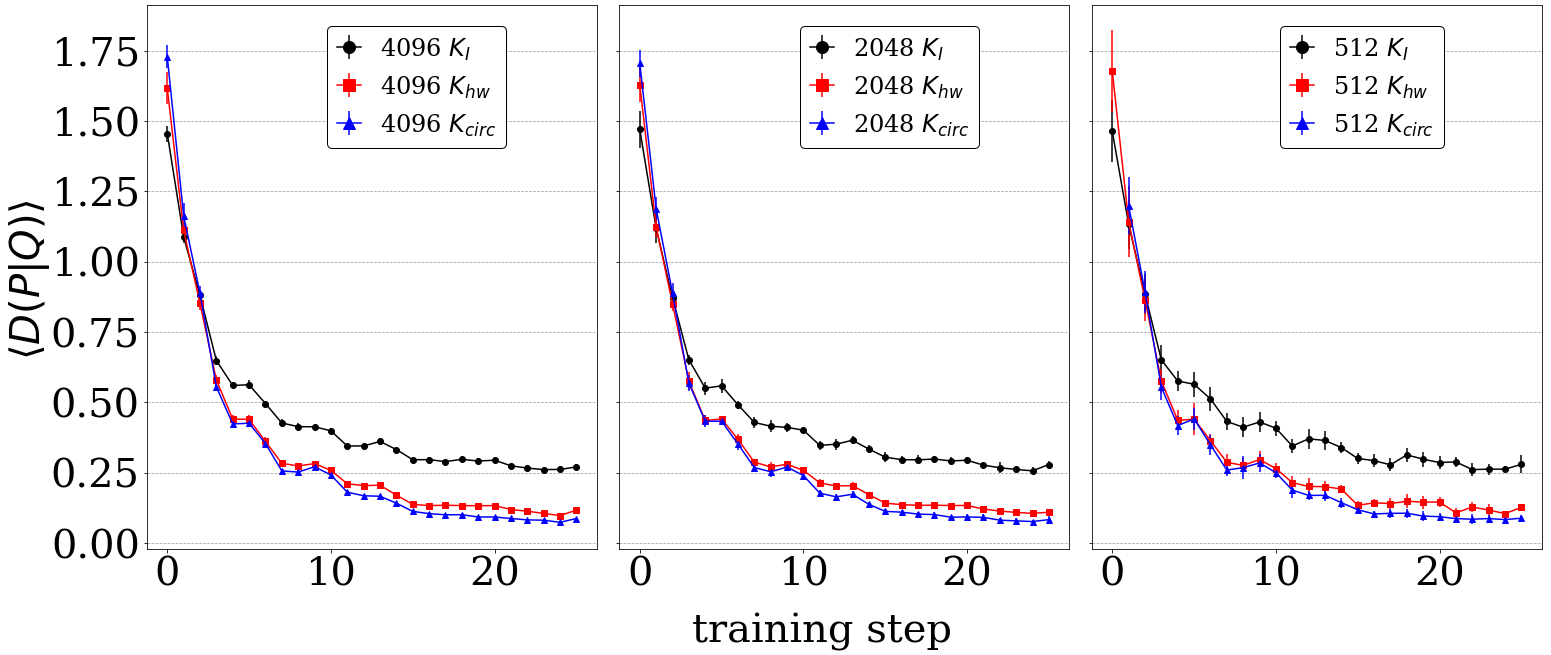}
  \vspace*{-0.4cm}
  \caption{$\langle D(P|Q) \rangle$ for $d_C=2$ circuits at $P_B$, mitigated with: $\mathbb{1}$ (black, circles), $\mathbb{K}_{hw}$ (red, square) and $\mathbb{K}_{circ}$ (blue, diamonds). Sub-sampled with: (Left) $4096$ shots, (Center) $2048$ shots, (Right) $512$ shots.}
  \label{fig:EMsubTVCP4}
\end{figure}

\begin{figure}[htbp]
  \includegraphics[width=\columnwidth]{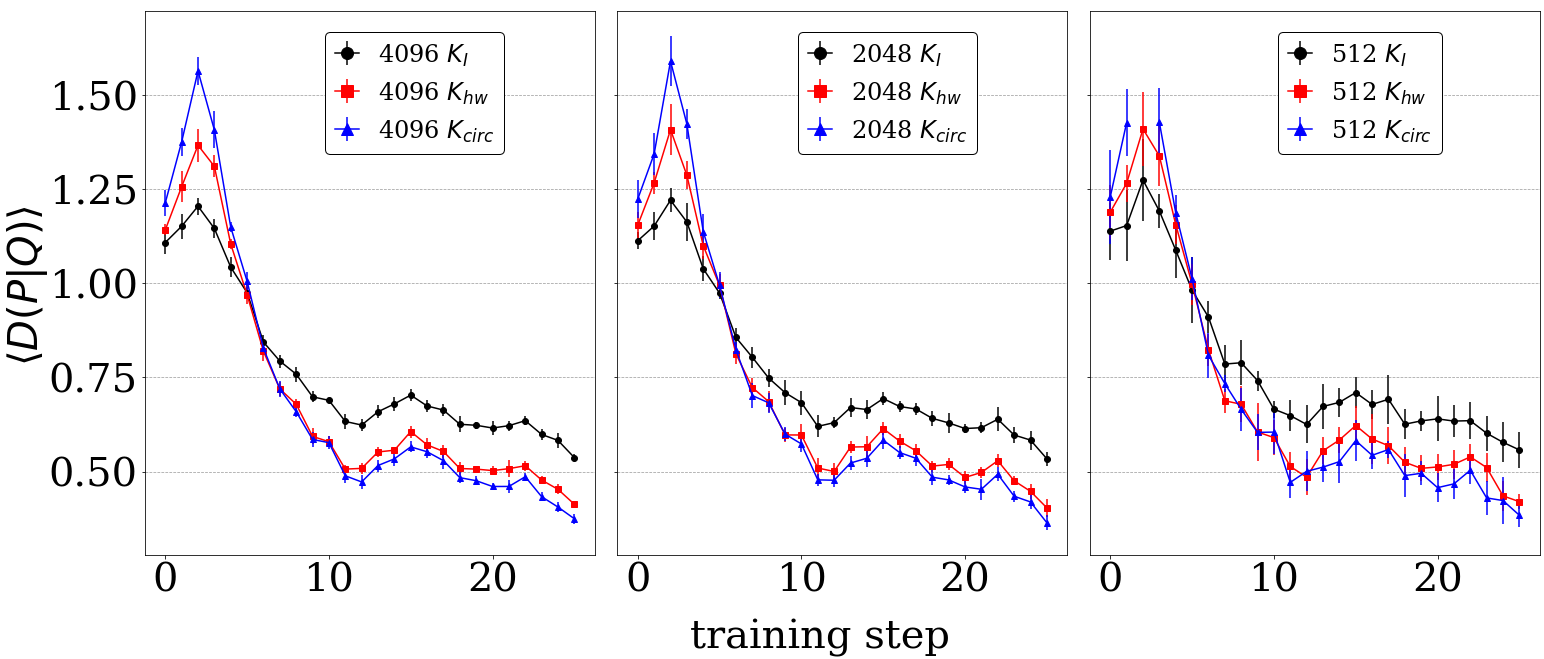}
  \vspace*{-0.4cm}
  \caption{$\langle D(P|Q) \rangle$ for $d_C=3$ circuits at $P_B$, mitigated with: $\mathbb{1}$ (black, circles), $\mathbb{K}_{hw}$ (red, square) and $\mathbb{K}_{circ}$ (blue, diamonds). Sub-sampled with: (Left) $4096$ shots, (Center) $2048$ shots, (Right) $512$ shots.}
  \label{fig:EMsubTCLP4}
\end{figure}

We plot average \KLmetric{} values ($\langle D(P|Q) \rangle$) defined by resampling a composite distribution $10$ times to show the reduction in $\langle D(P|Q) \rangle$ for single training runs at different hardware locations: in  \Cref{fig:EMsubTVCP4} we show the results for $d_C=2$ circuits and in \Cref{fig:EMsubTCLP4} we show the results for  $d_C=3$ circuits.  We plot $\langle D(P|Q) \rangle$ values to show that the reduction in $\langle D(P|Q) \rangle$ due to EM falls outside of the variance due to statistical noise. Error mitigation does not result in a reduction of $\langle D(P|Q) \rangle$ for all parameters: at early training parameters, error mitigation can lead to an increase or even divergence of \KLmetric{} is the amplitude of a BAS state in the mitigated distribution becomes zero.

\vspace*{-0.1cm}

In \Cref{tab:EM_Tokyo_KL_results} we summarize the lowest value $\langle D(P|Q) \rangle$ observed for the $d_C=2$ and $d_C=3$ circuits trained on \TOKYO{}  and post-processed with hardware- and circuit-based AEM.
 \begin{table}
      \centering
      \caption{$\min{\langle D(P|Q)\rangle}$ values observed in post-processed data for circuits trained on \TOKYO{}  sub-sampled with $2048$ shots.}
     \label{tab:EM_Tokyo_KL_results}

     \begin{tabular}{|l|c|c|c|c|}
     \hline
     \textbf{Circuit} & \textbf{Layout} & \textbf{$\mathbb{1}$ (post)} & \textbf{$\mathbb{K}_{hw}$ (post)} & \textbf{$\mathbb{K}_{circ}$ (post)} \\ \hline
     
$d_C=2$ & PA & 0.3333 & 0.1761 &  0.1509 \\ \hline
$d_C=2$ & PB & 0.2602 & 0.0988 &  0.0763 \\ \hline
$d_C=3$ & T1 & 0.4814 & 0.3142 &  0.2051 \\ \hline
$d_C=3$ & PB & 0.5390 & 0.4069 &  0.3750 \\ \hline
     \end{tabular}

 \end{table}

%% file: experiment1.tex
In this section we examine the effects of EM at each step in gradient-based circuit training.  The basic experiment remains the same as in \Cref{sec:experiment0}, except that rather than solely post-process the final distributions, we post-process at each training step and feed an error-mitigated loss function to the optimizer. The assignment error matrices are generated once at the start of training and used to mitigate the distributions evaluated throughout the training procedure.  To study the effect of error mitigation on the gradient-based training dynamics, we focus on the optimization of the \MMD{} function used to train the circuits and use this as a metric for EM efficacy in training. 

Queue times can lead to long time lags between AEM generation and when it is used in training, but the time between AEM generation and usage is shorter than in the previous section. On \BOBO{}: AEMs, initial circuit parameters, and distributions trained and generated between September 18--19 2019 and the longest time period between AEM generation and distribution evaluation was $21\mathrm{h}\;34\mathrm{min}$. On \VAL{}: AEMs, initial circuit parameters, and distributions were trained and generated between September 13--15 2019 and the longest time period between AEM generation and distribution evaluation was $16\mathrm{h}\;14\mathrm{min}$.  AEM were generated daily and the shortest time between AEM generation and final distribution evaluation for either \BOBO{} or \VAL{} was less than $75$ minutes.

\begin{figure}[htbp]
  \centering
  \includegraphics[width=\columnwidth]{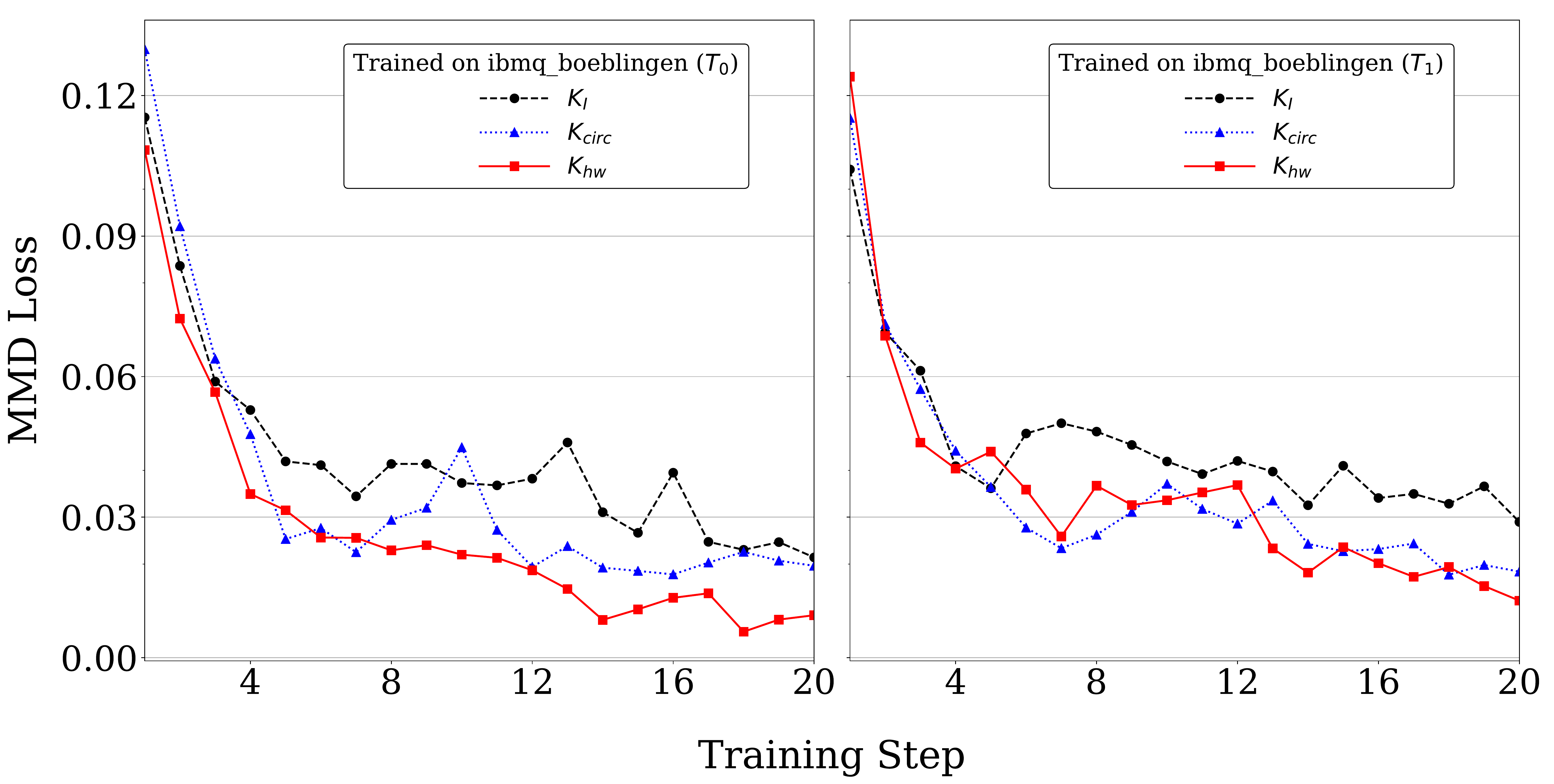}
  \caption{\MMD{} for a $d_C=3$ circuit during $20$ steps of training with Adam on \BOBO{} and $3$ AEM in training. (Left)Measured on $T_0$ for $\mathbb{K}_{\mathbb{1}}$ (black, circles, dotted), $\mathbb{K}_{hw}$ (red, squares, solid) and $\mathbb{K}_{circ}$ (blue, triangles, dashed). (Right) Measured on $T_1$ for $\mathbb{K}_{\mathbb{1}}$ (black, circles, dotted), $\mathbb{K}_{hw}$ (red, squares, solid) and $\mathbb{K}_{circ}$ (blue, triangles, dashed).}
  \label{fig:MMD_bobo_kernel_intrain}
\end{figure}

\begin{figure}[htbp]
  \centering
  \includegraphics[width=\columnwidth]{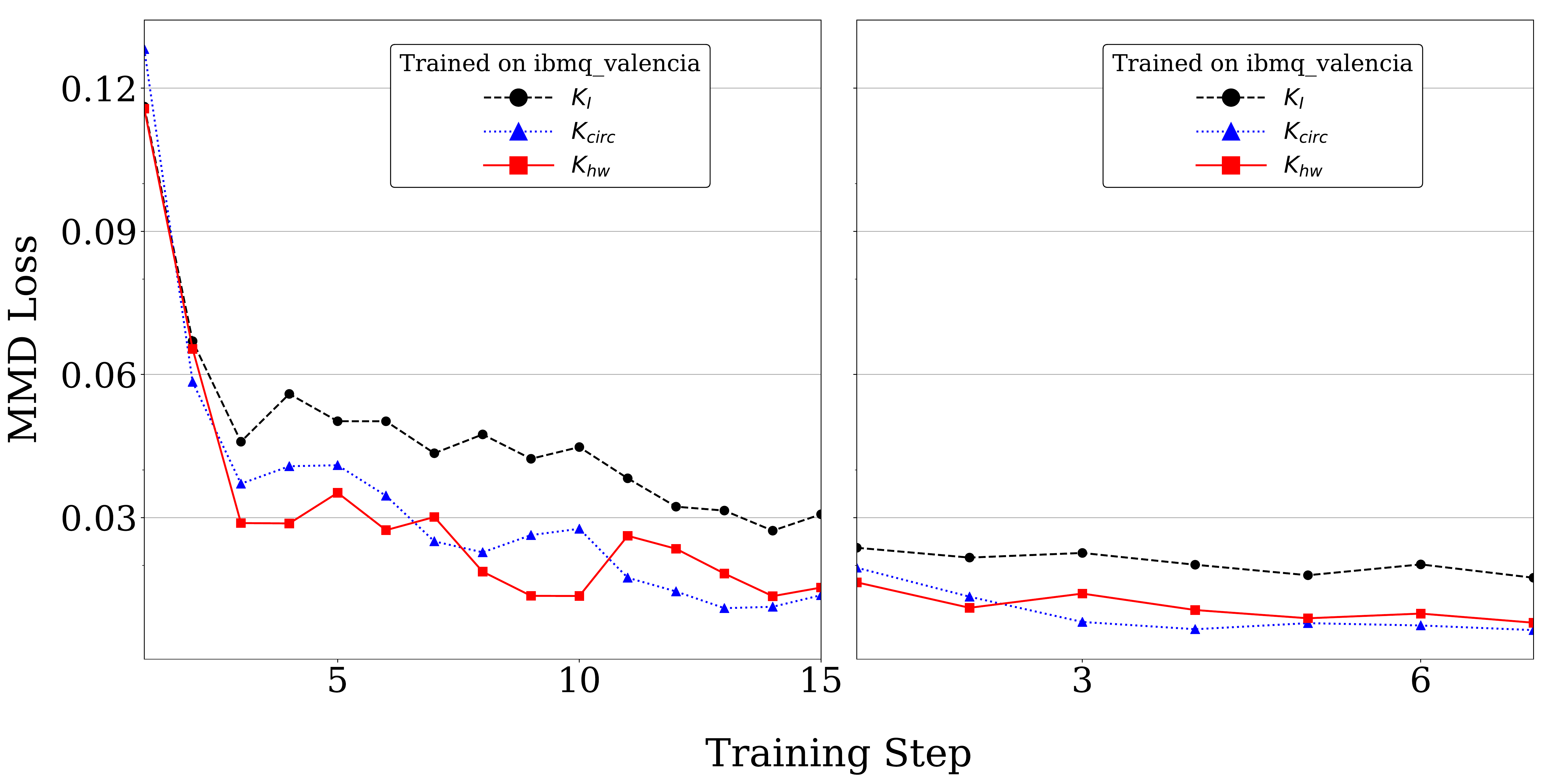}
  \caption{\MMD{} for a $d_C=3$ circuit measured during training with Adam on \VAL{} and $3$ AEM in training. (Left) $15$ steps of Adam training with $\mathbb{K}_{\mathbb{1}}$ (black, circles, dotted), $\mathbb{K}_{hw}$ (red, squares, solid) and $\mathbb{K}_{circ}$ (blue, triangles, dashed). (Right) $7$ steps of Adam from pre-trained values with $\mathbb{K}_{\mathbb{1}}$ (black, circles, dotted), $\mathbb{K}_{hw}$ (red, squares, solid) and $\mathbb{K}_{circ}$ (blue, triangles, dashed).}
  \label{fig:MMD_val_kernel_intrain}
\end{figure}

We trained the $d_C=3$ circuit for a fixed number of steps on different QPU available from IBM: \BOBO{} accessed via a cloud-based queue and \VAL{} accessed via a cloud-based reservation system.  It was verified that the compiled QASM circuit executed on each QPU used the minimum number of CNOT gates.  On \BOBO{} we trained each circuit for $20$ steps of Adam at two locations ($T_0,T_1$).  On \VAL{} we ran $2$ training sets: the first trained the circuit for $15$ steps of Adam starting from a random initialization; the second trained the circuit for $7$ steps of Adam starting from initial parameters that were pre-trained on \TOKYO{}.  In \Cref{fig:MMD_bobo_kernel_intrain,fig:MMD_val_kernel_intrain} we show \MMD{} evaluated during training using mitigated distributions and report the miminum MMD loss value measured during training in \Cref{tab:MMD_values_EMintrain}.  For this pre-trained circuit we reduced the learning rate to $\alpha = 0.1$.  

 \begin{table}
      \centering
      \caption{$\min{(\MMD)}$ observed in training $d_C=3$ circuits on \BOBO{} and \VAL{} with BAS(2,2) target.}
     \label{tab:MMD_values_EMintrain}

     \begin{tabular}{|p{2.5cm}|p{1.2cm}|p{1.2cm}|p{1.2cm}|}
     \hline
     \textbf{QPU (Layout)} & \textbf{$\mathbb{K}_\mathbb{1}$ (train) } & \textbf{$\mathbb{K}_{hw}$ (train)} & \textbf{$\mathbb{K}_{circ}$ (train)} \\ \hline
     
 \BOBO $\:$ ($T_0$) & 0.0214 & 0.00555 & 0.0178 \\ \hline
 \BOBO $\:$ ($T_1$) & 0.0290 & 0.0122 & 0.0177\\ \hline
 \VAL & 0.0273 & 0.0136 & 0.0110 \\ \hline
 \VAL $\:$ (pre-trained) & 0.0174 & 0.00799 & 0.00644 \\ \hline
     \end{tabular}

 \end{table}

\begin{figure}[htbp]
  \centering
  \includegraphics[width=\columnwidth]{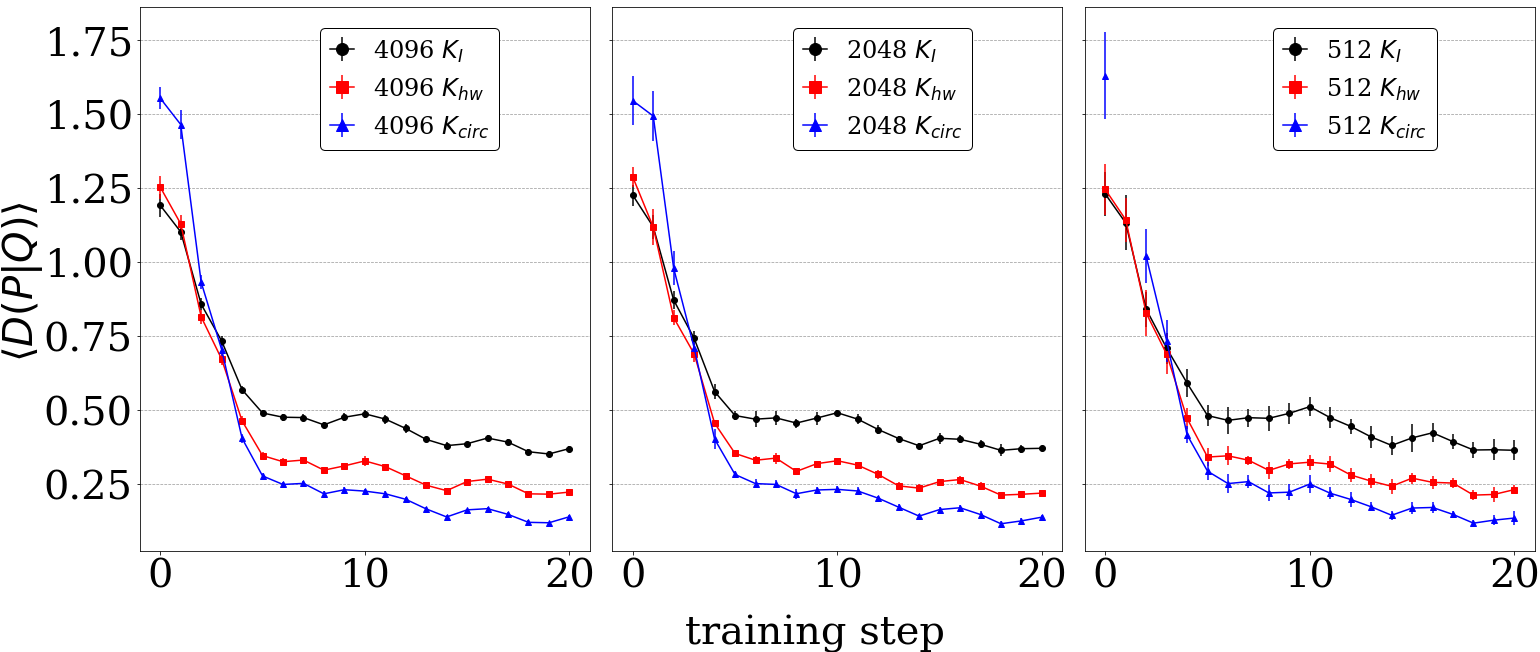}
  \caption{$\langle D(P|Q) \rangle$ for $d_C=3$ circuits trained with $\mathbb{K}_{hw}$ in training on \BOBO{} at $T_0$, sub-sampled with: (Left) 4096 shots, (Center) 2048 shots, (Right) 512 shots. The scores are post-processed with: $\mathbb{K}_{\mathbb{1}}$ (black, circles), $\mathbb{K}_{hw}$ (red, square) and  $\mathbb{K}_{circ}$ (blue, triangles).}
  \label{fig:KL_CL_T0_BOBO_EMtrain}
\end{figure} 
 
The circuit parameters recorded during training are used to generate multiple distributions at each step. For each set of parameters evaluated on \BOBO{}, a batch of $5$ identical circuits were sent to the hardware and evaluated with $2048$ shots to construct a composite distribution with effective shot size of $10240$.  For each set of parameters evaluated on \VAL, a batch of $2$ identical circuits were sent to the hardware and evaluated with 8192 shots.  These counts were used to construct a composite distribution with an effective shot size of $16384$. When evaluating the \KLmetric{} we apply post-processing as in \Cref{sec:experiment0}.  In \Cref{fig:KL_CL_T0_BOBO_EMtrain} we plot $\langle D(P|Q) \rangle$ for the circuit trained at $T_0$ with $\mathbb{K}_{hw}$.  This circuit had the lowest MMD values in training (see \Cref{fig:MMD_val_kernel_intrain}) and the reduction in $\langle D(P|Q) \rangle$ at later training steps is again far outside the variance induced by statistical noise. We report the minimum score values observed over the training data for all circuits in \Cref{tab:KL_CL_EMtrain_results}.

 \begin{table}
      \centering
      \caption{$\min{\langle D(P|Q)\rangle}$ values observed in post-processed data for $d_C=3$ circuits trained on \BOBO{} and \VAL.  Mean taken from $10$ distributions sub-sampled with $2048$ shots.}
     \label{tab:KL_CL_EMtrain_results}

     \begin{tabular}{|p{2.5cm}|p{1.cm}|p{1.cm}|p{1.cm}|p{1.cm}|}
     \hline
     \textbf{QPU (Layout)} & \textbf{AEM (train)} & \textbf{$\mathbb{K}_\mathbb{1}$
    (post)} & \textbf{$\mathbb{K}_{hw}$ (post)} & \textbf{$\mathbb{K}_{circ}$ (post)} \\ \hline

\BOBO{} ($T_0$) & $\mathbb{K}_\mathbb{1}$ & 0.3199 & 0.1827 & 0.1135 \\ \hline
\BOBO{} ($T_1$) & $\mathbb{K}_\mathbb{1}$ & 0.3701 & 0.2593 & 0.2426 \\ \hline
\BOBO{} ($T_0$) & $\mathbb{K}_{hw}$ & 0.3601 & 0.2169 & 0.1205 \\ \hline
\BOBO{} ($T_1$) & $\mathbb{K}_{hw}$ & 0.3345 & 0.2083 & 0.1950 \\ \hline
\BOBO{} ($T_0$) & $\mathbb{K}_{circ}$ & 0.4371 & 0.2831 & 0.2170 \\ \hline
\BOBO{} ($T_1$) & $\mathbb{K}_{circ}$ & 0.4060 & 0.2935 & 0.2808 \\ \hline\hline
\VAL & $\mathbb{K}_\mathbb{1}$ & 0.3726 & 0.2170 & 0.1776 \\ \hline
\VAL{} (pre-trained) & $\mathbb{K}_\mathbb{1}$ &  0.2938 & 0.1196 & 0.0869 \\ \hline
\VAL & $\mathbb{K}_{hw}$ & 0.3243 & 0.1668 & 0.1374 \\ \hline
\VAL{} (pre-trained) & $\mathbb{K}_{hw}$ &  0.2920 & 0.1261 & 0.0888 \\ \hline
\VAL & $\mathbb{K}_{circ}$ & 0.3560 & 0.2090 & 0.1770 \\ \hline
\VAL{} (pre-trained) & $\mathbb{K}_{circ}$ &  0.3115 & 0.1498 & 0.1133 \\ \hline
     \end{tabular}
 \end{table}

%% file: discussion.tex
Densely parameterized quantum circuits (circuits where the dimension of the gradient far exceeds the size of the qubit register) pose an interesting challenge to the problem of SPAM error mitigation.  We rely on the AEM to mitigate assignment errors (using $\mathbb{K}_{hw}$) and also general circuit noise (using $\mathbb{K}_{circ}$).  Our first point of discussion is how robust the post-processing is to the temporal characteristics of the hardware as captured by the AEM.  Our second point of discussion is how error mitigation in training is affected by the target distribution. 

\subsection{Temporal robustness of metric post-processing}
\label{sec:post_process_discuss}
Our approach to error mitigation relies on a matrix characterization of hardware noise. We do not attempt to characterize the noise of individual gates but instead gives an approximation of how SPAM errors manifest in preparing the individual basis states.  In \Cref{sec:experiment0} we showed that AEM kernels that were not generated at training time could be used to post-process distributions and obtain improved \KLmetric{} performance. This points to low hardware drift of qualitative noise properties over time. In \Cref{sec:experiment1}, when using error mitigation in our circuit training, the AEMs were generated at the start of training.  We access these quantumm devices remotely it may not be feasible to evaluate the AEM exactly at the time of training; for the case of long queue times the training steps can occur at significant time lags after generating a specific AEM.  Since it is known that the noise characteristics of quantum devices are time dependent, it is important to analyze the robustness of our approach to error mitigation with respect to time.

In the absence of any noise or error, $\mathbb{K}_{kernel} = \mathbb{1}$ and any basis state will be prepared with fidelity $1.0$.  To quantify the degree of ``noisiness'' in AEMs we calculate the distance between ($\mathbb{K}_{kernel}$) and the identity matrix using the Frobenius norm of ($\mathbb{1}-\mathbb{K}_{kernel}$):

\begin{equation}
 \| (\mathbb{1}-\mathbb{K}_{kernel}) \| = \sqrt{\sum_i \sum_j |(\delta_{ij}-\mathbb{K}_{kernel}(i,j))|^2}.
\end{equation}

In \Cref{tab:hw_frob} we report the individual norms of $\mathbb{1} -\mathbb{K}_{hw}$ and in \Cref{tab:circ_frob} we report the individual norms of $\mathbb{1} -\mathbb{K}_{circ}$ for multiple AEM that were generated over a $2$ week span.  For $\mathbb{K}_{hw}$, the norms have range $[0.803,1.25]$ and norms for $\mathbb{K}_{circ}$ generated over the same $2$ week span have range  $[1.03,1.59]$.  Fluctuations in hardware noise leads to significant differences between AEMs generated on different dates. 

\begin{table}
\centering
\caption{Frobenius norm $\| (\mathbb{1}-\mathbb{K}_{hw}) \|$ for AEM generated on \TOKYO.}
\label{tab:hw_frob}
\begin{tabular}{|p{2.0cm}|p{1.6cm}|p{1.6cm}|p{1.6cm}|}
\hline
Date & $P_A$ & $P_B$ & $T_1$ \\
\hline
07/25/2019 &  1.142 &  0.888 &  1.250 \\
\hline
07/30/2019 &  1.110 &  1.001 &  1.176 \\
\hline
07/31/2019 &  0.935 &  0.990 &  1.082 \\
\hline
08/03/2019 &  1.121 &  0.936 &  1.084 \\
\hline
08/04/2019 &  1.047 &  0.961 &  1.026 \\
\hline
08/06/2019 &  0.899 &  0.842 &  1.051 \\
\hline
08/07/2019 &  0.807 &  0.804 &  1.017 \\
\hline
08/08/2019 &  0.866 &  0.833 &  1.047 \\
\hline
\end{tabular}
\end{table}

For reference, the Frobenius norm of $\| (\mathbb{1}-\mathbb{K}_{hw}) \|$ for AEM generated on \TOKYO{} in May 2019 is $0.792$ ($P_A$), $0.836$ ($P_B$), and $0.923$ ($T_1$).  The Frobenius norm of $\| (\mathbb{1}-\mathbb{K}_{circ}) \|$ for AEM generated on \TOKYO{} in May 2019 is $1.746$ ($d_C=2$, $P_A$), $1.185$ ($d_C=2$, $P_B$), $1.207$ ($d_C=3$, $P_B$)and $1.425$ ($d_C=3$, $T_1$).  
The post-processed \KLmetric{} values are reported in \Cref{tab:EM_Tokyo_KL_results}, overall the $\mathbb{K}_{circ}$ led to the largest reductions in $\min{\langle D(P|Q)\rangle}$.  For $d_C=2$ circuits the \KLmetric{} was reduced by $54.7 \%$ (at $P_A$) and $70.6 \%$ (at $P_B$). For $d_C=3$ circuits the \KLmetric{} was reduced by $57.4 \%$ (at $T_1$), and $30.4 \%$ (at $P_B$).

\begin{table}
\centering
\caption{Frobenius norm $\| (\mathbb{1}-\mathbb{K}_{circ}) \|$ for AEM generated on \TOKYO.}
\label{tab:circ_frob}
\begin{tabular}{|p{2.0cm}|p{1.2cm}|p{1.2cm}|p{1.2cm}|p{1.2cm}|}
\hline
Date &  $d_C=2$ $(P_A)$ &  $d_C=2$ $(P_B)$ &  $d_C=3$ $(P_B)$ &  $d_C=3$ $(T_1)$ \\
\hline
07/25/2019 &        1.338 &        1.208 &        1.253 &        1.487 \\
\hline
07/30/2019 &        1.299 &        1.151 &        1.281 &        1.516 \\
\hline
07/31/2019 &        1.193 &        1.187 &        1.305 &        1.499 \\
\hline
08/03/2019 &        1.159 &        1.272 &        1.218 &        1.594 \\
\hline
08/04/2019 &        1.140 &        1.204 &        1.358 &        1.436 \\
\hline
08/06/2019 &        1.559 &        1.105 &        1.194 &        1.546 \\
\hline
08/07/2019 &        1.032 &        1.053 &        1.120 &        1.264 \\
\hline
08/08/2019 &        1.132 &        1.087 &        1.186 &        1.385 \\
\hline
\end{tabular}
\end{table}

\begin{figure}[htbp]
  \centering
  \includegraphics[width=\columnwidth]{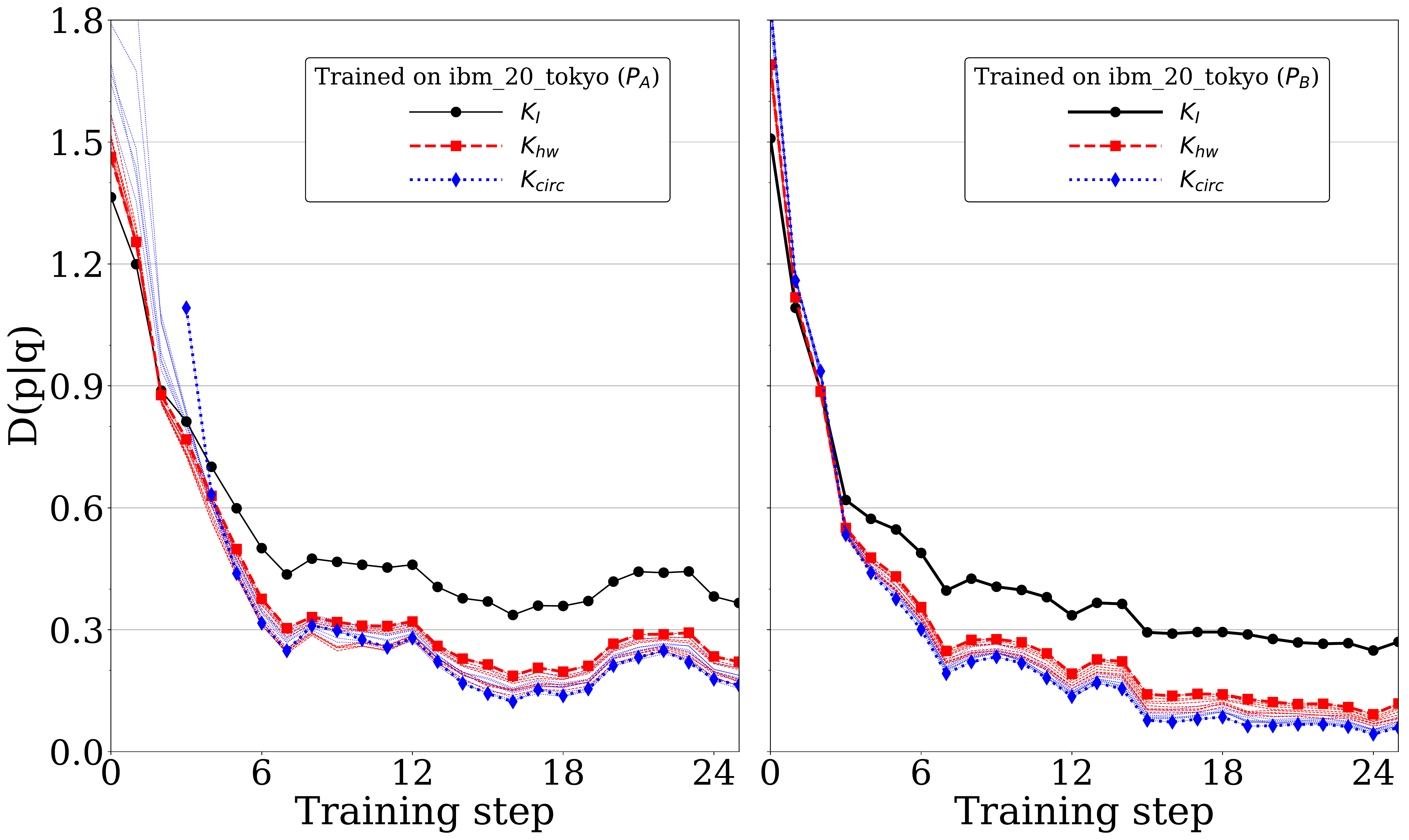}
  \caption{$D(P|Q)$ values for $d_C=2$ circuits trained on \TOKYO{} in May 2019, sampled at $2048$ shots, and post-processed with the $\mathbb{K}_{hw}$ (red, dashed) and $\mathbb{K}_{circ}$ (blue, dotted) summarized in \Cref{tab:hw_frob,tab:circ_frob}. Un-mitigated values (black, circles, solid) show as a reference. (Left) Circuits deployed on $P_A$, (Right) Circuits deployed on $P_B$.}
  \label{fig:VC_multi_kernel_plots}
\end{figure}

\begin{figure}[htbp]
  \centering
  \includegraphics[width=\columnwidth]{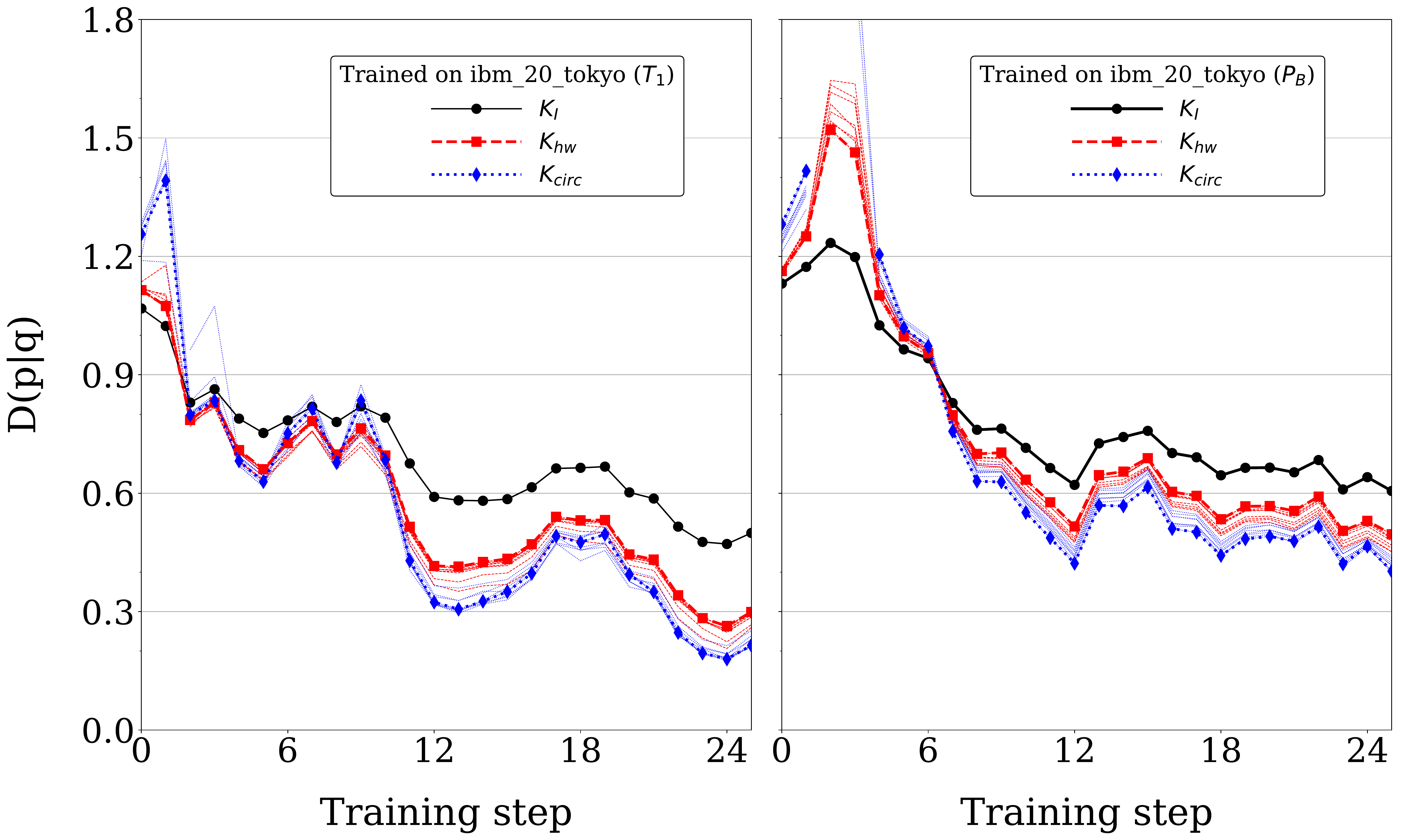}
  \caption{$D(P|Q)$ values for $d_C=3$ circuits trained on \TOKYO{} in May 2019, sampled at $2048$ shots, and post-processed with the $8$ $\mathbb{K}_{hw}$ (red, dashed) and $8$ $\mathbb{K}_{circ}$ (blue, dotted) summarized in \Cref{tab:hw_frob,tab:circ_frob}. Un-processed values (black, circles, solid) show as a reference. (Left) Circuits deployed on ($T_1)$, (Right) circuits deployed on ($P_B$).}
  \label{fig:CL_multi_kernel_plots}
\end{figure}

Using the set of AEMs generated in July--August $2019$, we re-ran the post-processing methods of \Cref{sec:experiment0} on circuits trained on \TOKYO{} and plot the values in \Cref{fig:VC_multi_kernel_plots,fig:CL_multi_kernel_plots}.  When these AEMs are used to post-process the same data as in \Cref{sec:experiment0}, we see similar reductions in the \KLmetric{}.  Again the $\mathbb{K}_{circ}$ again gave maximum reductions in the $\min{\langle D(P|Q) \rangle}$ value.  For $d_C=2$ circuits the maximum reduction in the \KLmetric{} was $63.5 \%$ at $P_A$ and $82.6 \%$ at $P_B$. For the $d_C=3$ circuits the maximum reduction in the \KLmetric{} value was $62.7 \%$ at $T_1$ and $33.8 \%$ at $P_B$.  

The BAS(2,2) distribution (\Cref{fig:BAS}) should be difficult to prepare due to its high entanglement entropy \cite{benedetti2019generative}, but in order to test if something specific about this target distribution makes EM in post-processing particularly effective, we expanded the benchmark with additional experiments test for edge cases. For example, after several steps of training the amplitudes of the BAS states are greater than the non-BAS states; this distribution can be ``corrected`` using a simple threshold function, $f(x) = \Theta(p(x)-p_0)$.  Applying $f(x)$ to the final sampled state of a circuit would suppress states with probabilities $p(x)<p_0$.  If the amplitude of all BAS states are above the threshold value
then only non-BAS states are removed and the KL score will improve with no assumption about the state being spurious or not. To test if our EM method is oversimplifying the distribution in the manner outlined above, we applied it to other distributions for which $f(x)$ would fail at improving fidelity to the target distribution, which we outline in the next section.

\subsection{Robustness of gradient-based training}
\label{sec:mitigated_training_discussion}
When error mitigation (see \Cref{sec:methods}) is incorporated into gradient-based training the classical optimizer updates the circuit parameters from information in the mitigated loss gradient.  Post-processing discards counts (i.e. information) from the final sampled state that is assumed to be spurious, however we need to ensure that information relevant to state learning is not lost.  

In \Cref{sec:experiment0,sec:experiment1} all circuits were trained with the BAS(2,2) distribution as a target.  The results presented in \Cref{sec:experiment0} show that using mitigated distributions to evaluate the \KLmetric{} can improve performance over un-mitigated distributions, but in \Cref{sec:post_process_discuss} we conjectured that the improvements in the \KLmetric{} and the robustness of the metric post-processing might be due to EM simply discarding low amplitude states indiscriminately (which would be deleterious to the metric for more complex distributions that contain eigenstates with low probability amplitudes).  If that was the case, then this would be reflected in the training dynamics and specifically in the optimization of the MMD loss function.  Yet in \Cref{sec:experiment1} we observe that the MMD values recorded during training show that the circuits with AEM mitigation tend to reach lower loss values compared to the non-mitigated $\mathbb{K}_{\mathbb{1}}$ circuits (see  \Cref{fig:MMD_bobo_kernel_intrain,fig:MMD_val_kernel_intrain} ).

To further explore the effects of error mitigation in training, we repeated the same experiments in \Cref{sec:experiment1} on \VAL{} using a second set of distributions, shown in \Cref{fig:BAS}.  In contrast to the BAS(2,2) distribution (\Cref{fig:BAS}), training a circuit to fit these new distributions requires that low-amplitude states are learned and not discarded. The highest weight state $|1111\rangle$ has zero amplitude in Distribution $1$ and the lowest weight state $|0000\rangle$ has zero amplitude in Distribution $2$.  The smallest non-zero amplitude in each distribution was $p(x_i) = 0.0005$ and training circuits using $2048$ shots gives sufficient resolution.  

It would be difficult to learn the target distribution with high accuracy at this shot size, but our goal was to observe qualitative behavior of the AEM in training. In \Cref{tab:MMD_Poissons} we summarize the lowest MMD values measured during training and in \Cref{tab:KL_scores_Poisson} we summarize the lowest $\langle \KL{P}{Q}\rangle$ scores found using post-processing.  Comparing the MMD values from \Cref{tab:MMD_Poissons} to the \KLmetric{} values with $\mathbb{K}_\mathbb{1}$ in post we see that the inclusion of EM in training led to lower values of both the MMD loss and \KLmetric{}.  However the inclusion of EM in post-processing led to interesting behavior.  

In \Cref{sec:experiment0} we observed that post-processing with $\mathbb{K}_{circ}$ returned the lowest \KLmetric{} values for the BAS(2,2) distribution, and in \Cref{sec:experiment1} circuits with EM in training and EM in post-processing had lower \KLmetric{} values.  In \Cref{tab:KL_scores_Poisson} we observe that the circuit with $\mathbb{K}_{circ}$ in training and $\mathbb{K}_{circ}$ in post-processing increased the \KLmetric{} value; whereas the circuit trained with $\mathbb{K}_{hw}$ in training and $\mathbb{K}_{hw}$ in post-processing returned the lowest \KLmetric{} value.  Since Distributions $1$ and $2$ contain very low amplitude states we re-evaluated the \KLmetric{} values with a high shot size (\Cref{tab:KL_scores_Poisson4096}), yet the qualitative behavior remained the same.  For these target distributions post-processing frequently causes the low-amplitude states to have zero amplitude in the mitigated distribution and the \KLmetric{} is infinite at several training steps.

\begin{figure}[htbp]
  \centering
  \includegraphics[width=\columnwidth]{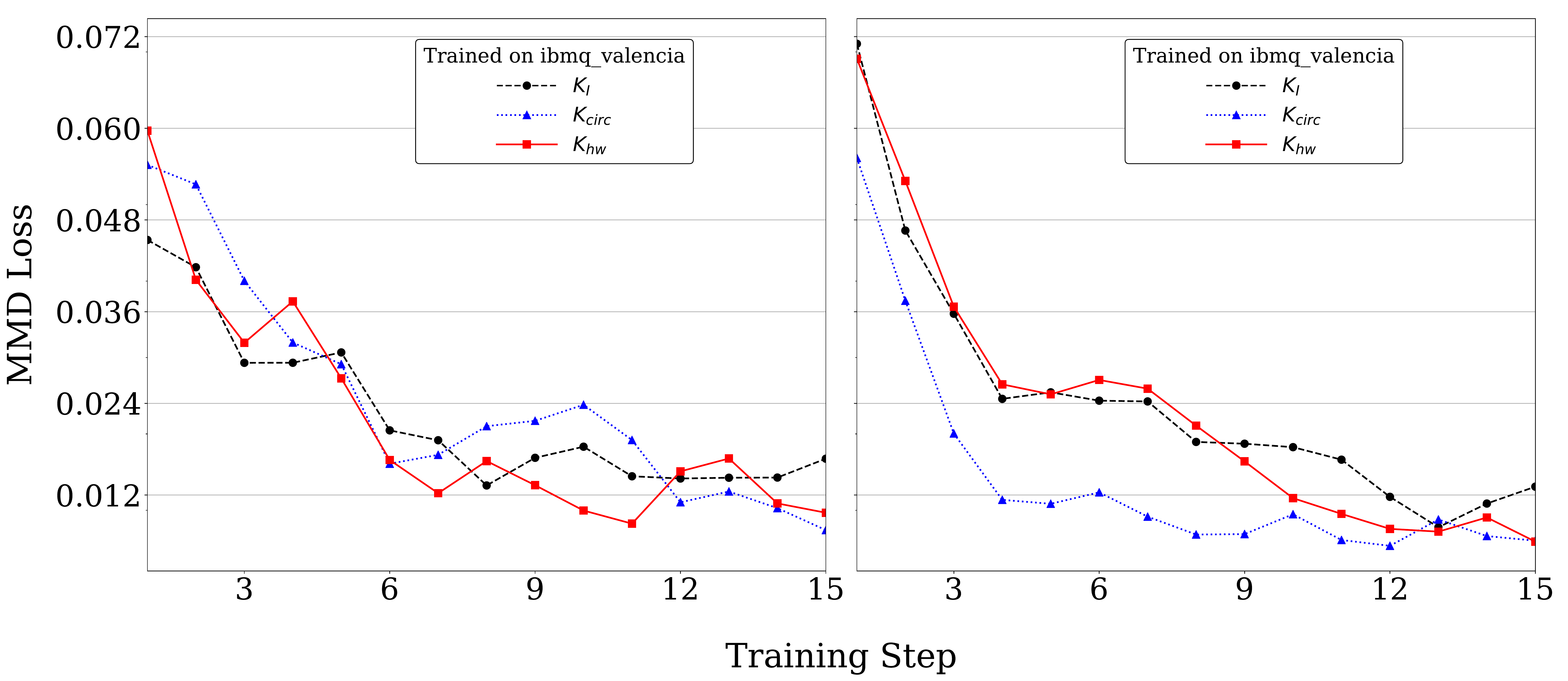}
  \caption{$\mathcal{L}_{MMD}$ measured on \VAL{} during $15$ steps of training with Adam using $3$ AEM in training: $\mathbb{K}_{\mathbb{1}}$ (black, circles, dotted), $\mathbb{K}_{hw}$ (red, squares, solid) and $\mathbb{K}_{circ}$ (blue, diamonds, dashed).   (Left) Target is Distribution 1 (right) Target is Distribution $2$.}
  \label{fig:MMD_Poissons}
\end{figure} 
 \begin{table}
      \centering
      \caption{$\min{(\mathcal{L}_{MMD})}$ values observed in training on \VAL.}
     \label{tab:MMD_Poissons}

     \begin{tabular}{|l|c |c|c|}
     \hline
     \textbf{Target} & \textbf{$\mathbb{K}_\mathbb{1}$ (train) } & \textbf{$\mathbb{K}_{hw}$ (train)} & \textbf{$\mathbb{K}_{circ}$ (train)}  \\ \hline
     
Distribution 1& 0.0133 & 0.00826 & 0.00741 \\ \hline
Distribution 2& 0.00778 & 0.00591 & 0.00535\\ \hline
     \end{tabular}

 \end{table}
 \begin{table}
      \centering
      \caption{$\min{\langle D(P|Q)\rangle}$ values observed in post-processed data for $d_C=3$ circuits trained on \VAL.  Mean taken from $10$ distributions sub-sampled with $2048$ shots.}
     \label{tab:KL_scores_Poisson}

     \begin{tabular}{|p{2.5cm}|p{1.2cm}|p{1.2cm}|p{1.2cm}|p{1.2cm}|}
     \hline
     \textbf{Target} & \textbf{AEM (train)} & \textbf{$\mathbb{K}_\mathbb{1}$ (post) } & \textbf{$\mathbb{K}_{hw}$ (post)} & \textbf{$\mathbb{K}_{circ}$ (post)}  \\ \hline
     
Distribution 1&$\mathbb{K}_\mathbb{1}$ &0.1481 & 0.1174 & 0.1724 \\ \hline
Distribution 1&$\mathbb{K}_{hw}$ &0.1423 & 0.1088 & 0.1229 \\ \hline
Distribution 1&$\mathbb{K}_{circ}$ &0.1200 & 0.1190 & 0.2783 \\ \hline\hline
Distribution 2&$\mathbb{K}_\mathbb{1}$ &0.1807 & 0.1954 & 0.2093\\ \hline
Distribution 2&$\mathbb{K}_{hw}$ &0.1230 & 0.1049 & 0.1013\\ \hline
Distribution 2&$\mathbb{K}_{circ}$&0.1215 & 0.0923 & 0.0883\\ \hline
     \end{tabular}

 \end{table}

 \begin{table}
      \centering
      \caption{$\min{\langle D(P|Q)\rangle}$ values observed in post-processed data for $d_C=3$ circuits trained on \VAL.  Mean taken from $10$ distributions sub-sampled with $4096$ shots.}
     \label{tab:KL_scores_Poisson4096}

     \begin{tabular}{|p{2.5cm}|p{1.2cm}|p{1.2cm}|p{1.2cm}|p{1.2cm}|}
     \hline
     \textbf{Target} & \textbf{AEM (train)} & \textbf{$\mathbb{K}_\mathbb{1}$ (post) } & \textbf{$\mathbb{K}_{hw}$ (post)} & \textbf{$\mathbb{K}_{circ}$ (post)}  \\ \hline
     
Distribution 1&$\mathbb{K}_\mathbb{1}$ & 0.1427 & 0.1175 & 0.1698 \\ \hline
Distribution 1&$\mathbb{K}_{hw}$ & 0.1372 & 0.1085 & 0.1211 \\ \hline
Distribution 1&$\mathbb{K}_{circ}$ & 0.1169 & 0.1264 & 0.1747 \\ \hline\hline
Distribution 2&$\mathbb{K}_\mathbb{1}$ & 0.1836 & 0.1962 & 0.1960\\ \hline
Distribution 2&$\mathbb{K}_{hw}$ & 0.1246 & 0.1074 & 0.0988\\ \hline
Distribution 2&$\mathbb{K}_{circ}$& 0.1170 & 0.0953 & 0.0856\\ \hline
     \end{tabular}

 \end{table}

In \Cref{sec:experiment1} we can observe qualitative effects of incorporating error mitigation from the overall behavior of the MMD loss function during training.  If the error mitigation step was arbitrarily discarding low count states then we would expect to see a severe reduction in the MMD minimization. While the addition of error mitigation inside the gradient-based training does not appear to result in catastrophic loss of information there are some subtle effects with different target distributions.  With the Bars and Stripes target, the inclusion of error mitigation inside the gradient-based training led to lower MMD values earlier in training for most circuits- the only exception was the circuit trained on \BOBO{} with $\mathbb{K}_{circ}$ at $T_0$ (see \Cref{fig:MMD_bobo_kernel_intrain}).  When the same circuit ansatz was trained on \VAL{} the non-trivial AEM resulted in lower MMD values.  When we trained QCBMs with respect to non-uniform distributions we notice that overall the lowest MMD values are observed when non-trivial AEMs are used in training.  While a significant speedup in training may be observed, (as seen in \Cref{fig:MMD_Poissons} for $\mathbb{K}_{circ}$ and Distribution $2$) the training dynamics with the trivial AEM can also be very similar to the training dynamics with non-trivial AEM (as seen in \Cref{fig:MMD_Poissons} for Distribution $1$).   These results are promising but suggest that the performance of DDCL is dependent on the choice of optimizer, target distribution.  Combined with the AEM asymmetry means a more in-depth study is needed to investigate the efficacy of error mitigation for gradient-based training.

%% file: conclusions.tex
In this work we consider two challenges faced when training parameterized circuits on NISQ devices:  how to effectively optimize the parameters of a circuit in the presence of noisy sampling, and how to extract a baseline measure of metrics from noisy evaluations.  Observations made in previous studies, such as the need to optimize circuit design for sparsity and hardware layout, still play a significant role even with the addition of error mitigation.  The \KLmetric{} gives us the final circuit performance and we use this to explore the efficacy of error mitigation in post-processing.  

Without any error mitigation in either the gradient-based training, or the evaluation of the \KLtext{} metric, the lowest \KLmetric{} value we observed for QCBMs trained for $25$ steps of Adam on noisy hardware was $0.2602$. This was observed for the sparsest circuit class $d_C=2$ optimized to run on \TOKYO. A recent study of QCBM training on Rigetti hardware have reported \KLtext{} values for trained QCBMs on the order of $0.1$ \cite{leyton2019robust}.  We set a threshold on the \KLtext{} metric of $\langle \KL{p}{q} \rangle < 0.1$ then we find that with error mitigation, two circuits can reach this threshold for the BAS(2,2) distribution.  The $d_C=2$ circuit trained on \TOKYO{} reached this threshold with error mitigation in metric post-processing ($\langle \KL{p}{q} \rangle =0.0763$), and the $d_C=3$ circuit trained on \VAL{} with error mitigation in training and also post-processing ($\langle \KL{p}{q} \rangle =0.0832$ with random initialization, $\langle \KL{p}{q} \rangle =0.0888$ with pre-training).  For the second set of distributions, the $d_C=3$ circuit was able to surpass this threshold for Distribution $2$ ($\langle \KL{p}{q} \rangle =0.0883$).  In \Cref{sec:experiment0} post-processing a measured distribution with an AEM reduced the value of the \KLmetric{} divergence by reducing the amplitude of states that are assumed to be spurious (the result of assignment error).  However this assumption can cause divergence of the \KLmetric{} which was observed for the Bars and Stripes distribution at parameters saved early in training, where it is possible for a BAS state to have low amplitude, and in general for target distributions that contains low amplitude states (Distribution $1,2$).

The performance of the MMD loss function over multiple training runs with EM included shows that EM typically causes a decrease in loss function at earlier training steps, as in Fig.~\ref{fig:MMD_Poissons} for Distribution 2, and in Figs.~\ref{fig:MMD_bobo_kernel_intrain} and \ref{fig:MMD_val_kernel_intrain} for the BAS distribution. This means that EM can decrease training time. Thus, a usable metric when studying loss functions with EM in training would be time to convergence, or resources required to achieve the minimal value. In several cases, we observe that training with EM qualitatively improves this metric.

In this work we studied how to benchmark the efficacy of gradient-based optimization in the presence of device noise with the addition of matrix-based method error mitigation. The training of parametrized circuits is a complex computational task that is dependent on the circuit ansatz, target distribution, classical optimization method and the time-dependence of device noise.    
Nonetheless we have observed benefits to using matrix-based error mitigation. In the future, the benchmark will be improved to include additional distributions, and as pulse-level controls become available to the general user, lower level circuit designs. The mitigation of hardware noise is also a complex problem and it may be necessary to supplement AEM with other error mitigation techniques (e.g. Richardson extrapolation \cite{temme2017error,li2017efficient,havlivcek2019supervised}, probabilistic cancellation \cite{temme2017error}, pulse stretching \cite{Kandala2019} or modified circuits \cite{endo2018practical}).

%% file: appendix.tex
\section{Mitigation SWAP gate noise}
\label{appendix:SWAP_noise}
The motivation behind the design of circuit-specific assignment error matrices is to attempt to capture additional noise and errors outside of the measurement error.  In \Cref{sec:experiment0,sec:experiment1,sec:discussion} we saw that using the circuit-specific AEM to post-process a measured distribution led to lower values of the \KLtext{} metric compared to the more general, hardware AEM and lower MMD values during training.  

In the main text we discussed circuits with sparse entangling layers, trained on hardware with connectivity that supports the gate layout.  The $d_C=3$ circuit was trained on $3$ qubit devices (\BOBO, \VAL, \TOKYO) but the $d_C=2$ circuit was only trained on \TOKYO.  The sparsity of the entangler layers and the optimization with hardware layout is designed to reduce the overall amount of two-qubit gate noise by reducing the number of noisy gates in a compiled circuit. The compiled circuits executed on hardware in the main text had the minimum number of CNOTs ($4$ and $6$, for $d_C=2$ and $d_C=3$ respectively).  

If a circuit is executed on hardware that does not exactly match the hardware coupler layout, then additional CNOTs will be added to the compiled circuit.  We tested the efficacy of error mitigation for noise associated with added CNOTs by running the same experiments from \Cref{sec:experiment1} using a $d_C=2$ circuit on \BOBO.  On \TOKYO{} the compiled circuit contained only $4$ CNOT gates, but on \BOBO{} the compiled circuit could have up to $16$ CNOT gates.

 \begin{figure}[htbp]
  \centering
  \includegraphics[width=\columnwidth]{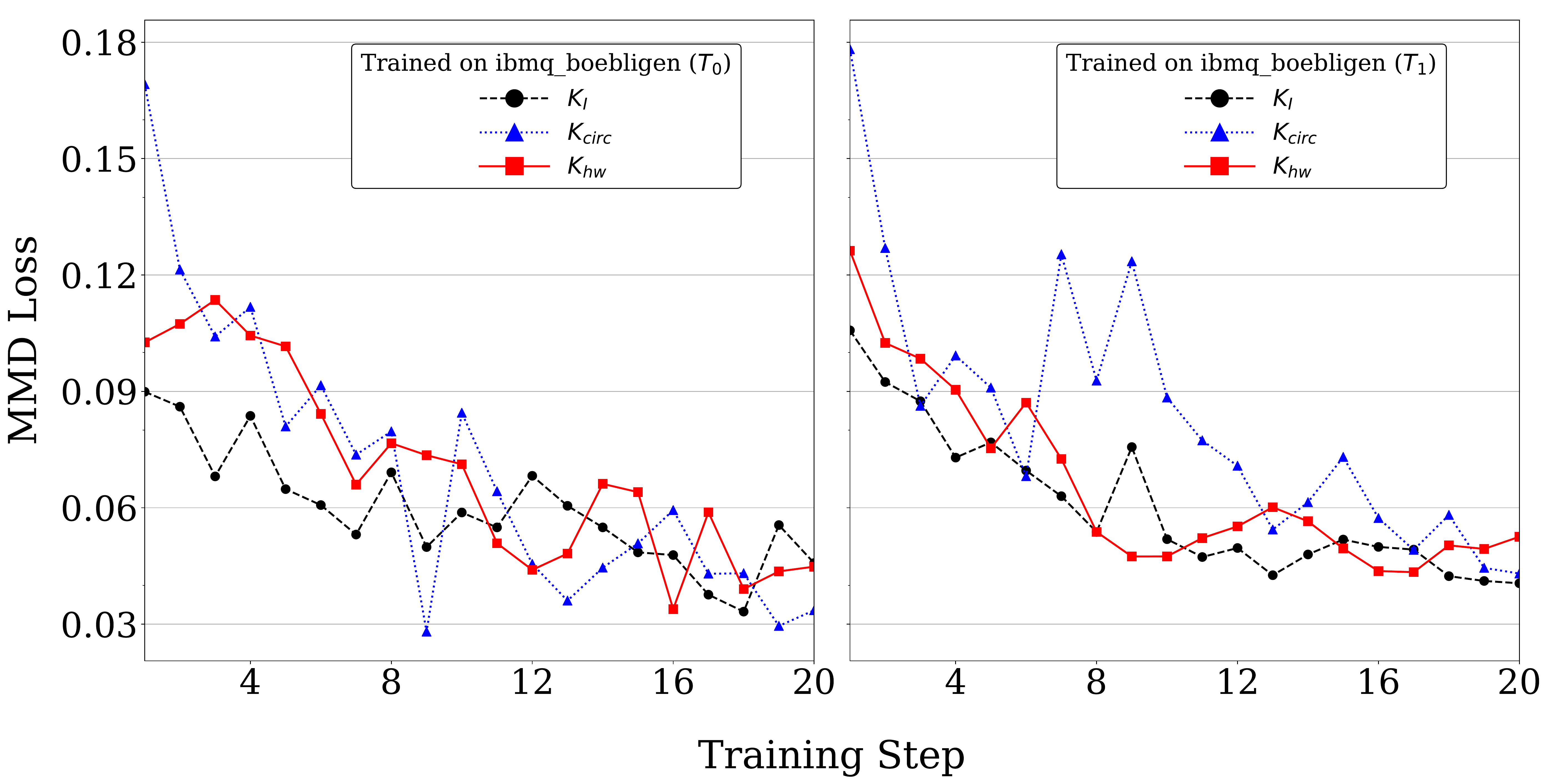}
  \caption{
  \MMD{} for a $d_C=2$ circuit during $20$ steps of training with Adam on \BOBO{} and $3$ AEM in training. (Left)Measured on $T_0$ for $\mathbb{K}_{\mathbb{1}}$ (black, circles, dotted), $\mathbb{K}_{hw}$ (red, squares, solid) and $\mathbb{K}_{circ}$ (blue, diamonds, dashed). (Right) Measured on $T_1$ for $\mathbb{K}_{\mathbb{1}}$ (black, circles, dotted), $\mathbb{K}_{hw}$ (red, squares, solid) and $\mathbb{K}_{circ}$ (blue, diamonds, dashed).}
  \label{fig:BAS_BOBO_VC_MMD}
\end{figure} 

Now, circuits trained with $\mathbb{K}_{hw}$ or $\mathbb{K}_{circ}$ in training shows no substantial improvement in the MMD loss minimization, and in \Cref{fig:BAS_BOBO_VC_MMD} we see that the inclusion of EM in training can impair the minimization of MMD over training.  The $\min{(\mathcal{L}_{MMD})}$ measured was $0.0332$, $0.033$, and $0.0281$ for $\mathbb{1}$,$\mathbb{K}_{hw}$ and $\mathbb{K}_{circ}$, respectively.  Yet post-processing did show improvement in the \KLtext{} metric (reported in \Cref{tab:KL_VC_EMtrain_results}).  In the main text we observed a lowest value of $\min{\langle D(P|Q) \rangle}= 0.0763$  for a $d_C=2$ circuit trained on \TOKYO.  Now on \BOBO{} with additional CNOTs the lowest value of $\min{\langle D(P|Q) \rangle}= 0.1490$.

 \begin{figure}[htbp]
  \centering
  \includegraphics[width=\columnwidth]{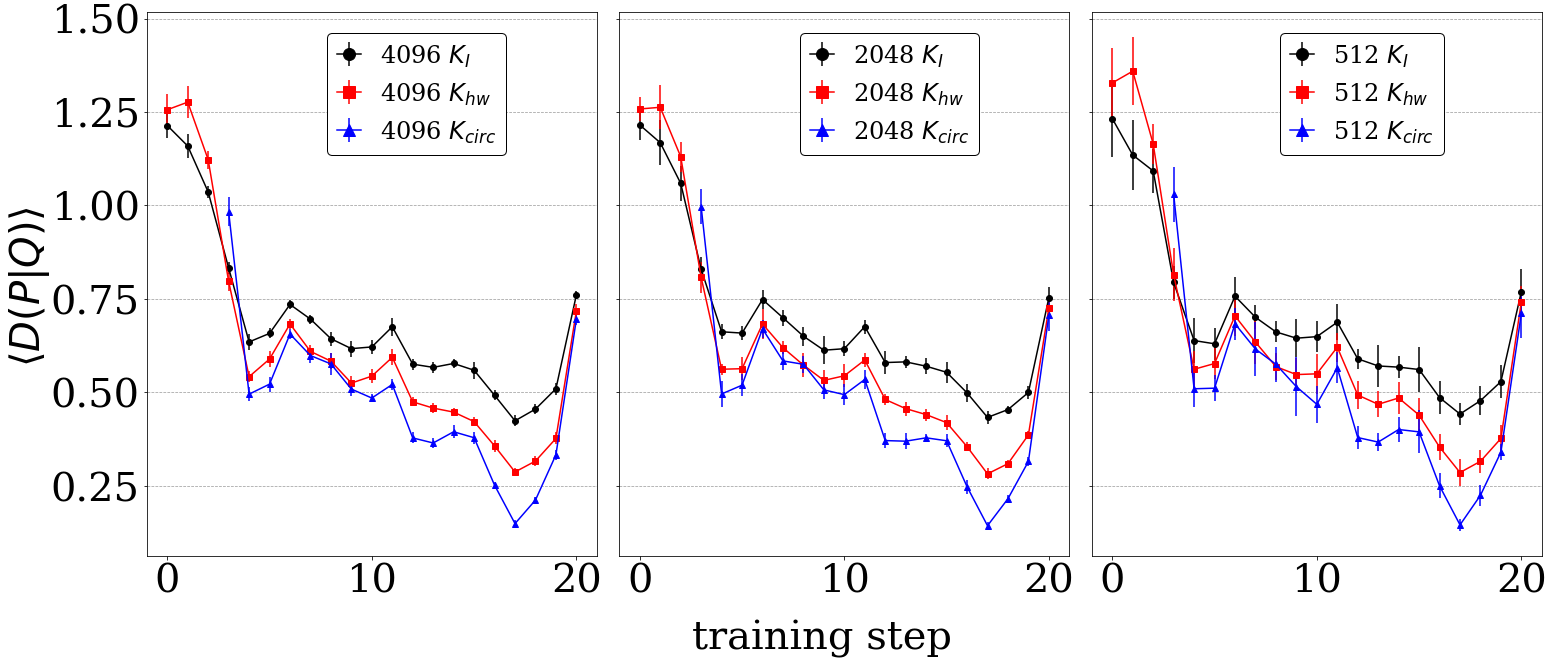}
  \caption{$\langle D(P|Q) \rangle$ for $d_C=2$ circuits with identity-AEM in training on \BOBO{}  random initialization. Averaged over $10$ sub-samples of (Left) 4096 shots, (Center) 2048 shots, (Right) 512 shots. The scores are post-processed with: no error mitigation (black, circles), hardware-AEM mitigation (red, square) and circuit-AEM mitigation (blue, diamonds).}
  \label{fig:EMsubs}
\end{figure} 
 
 \begin{table}
\centering
\caption{$\min{\langle D(P|Q)\rangle}$ values observed in post-processed data for $d_C=2$ circuits trained on \BOBO.  Mean taken from $10$ sub-sampled distributions of $2048$ shots.}
\label{tab:KL_VC_EMtrain_results}

     \begin{tabular}{|p{2.5cm}|p{1.2cm}|p{1.2cm}|p{1.2cm}|p{1.2cm}|}
     \hline
     \textbf{QPU (Layout)} & \textbf{AEM (train)} & \textbf{$\mathbb{1}$ (post)} & \textbf{$\mathbb{K}_{hw}$ (post)} & \textbf{$\mathbb{K}_{circ}$ (post)} \\ \hline

\BOBO{} ($T_0$) & $\mathbb{1}$ & 0.4373 & 0.2807 & 0.1490 \\ \hline
\BOBO{} ($T_0$) & $\mathbb{K}_{hw}$ & 0.5735 & 0.4779 & 0.3920 \\ \hline
\BOBO{} ($T_0$) & $\mathbb{K}_{circ}$ & 0.5967 & 0.5042 & 0.3990 \\ \hline\hline
\BOBO{} ($T_1$) & $\mathbb{1}$ & 0.4632 & 0.4022 & 0.2552 \\ \hline
\BOBO{} ($T_1$) & $\mathbb{K}_{hw}$ & 0.5827 & 0.5154 & 0.4314 \\ \hline
\BOBO{} ($T_1$) & $\mathbb{K}_{circ}$ & 0.6649 & 0.6097 & 0.5124 \\ \hline
\end{tabular}
\end{table}
\section{Assignment error matrix construction}
\label{appendix:AEM}
Our approach to error mitigation is based on the concept of assignment error matrices and we extract the error correlations from the output of multi-qubit circuits executed on localized qubit subsets.  For a system of $N$ qubits, a set of $2^N$ quantum circuits is used to prepare each individual state of the computational basis.  The general hardware-AEM only mitigates readout errors and is generated using  shallow circuits of $N$ qubits and up to $N$ individual $X$ gates.  We extend this AEM concept to a more general structure so we can estimate the general error associated with a particular parameterized circuit.  To construct the circuit-AEM we use the parameterized circuit ansatz shown in \Cref{fig:ansatz_entanglers} to prepare the $2^N$ basis states.  

Since we project the final state prepared by the circuit onto a set of real-valued amplitudes there are multiple sets of rotational parameters that will result in the same final distribution.  We constructed the matrices $\mathbb{K}_{circ}$ used in the main text using the general circuit ansatz of \Cref{fig:ansatz_entanglers}.  The rotational parameters in the first and final rotation layers are all set to zero.  Each qubit in the middle rotation layer has a gate sequence $R_X(\theta_i)R_Z(\theta_j)R_X(\theta_k)$ applied to it. By replacing this gate sequence with either $R_X(0)R_Z(0)R_X(0)$ or $R_X(-\pi/2)R_Z(-\pi)R_X(-\pi/2)$ we can use the parameterized ansatz to prepare any basis state.  For example, applying $R_X(0)R_Z(0)R_X(0)$ to all $4$ qubits will return only the state $|0000 \rangle$ in the absence of noise.  

\begin{table}
\centering
\caption{Individual state fidelities for $d_C=2$, $d_C=3$ $\mathbb{K}_{circ}$ and $\mathbb{K}_{hw}$ all evaluated at $P_B$ on \TOKYO.}
\label{tab:state_fidelities}
\begin{tabular}{| p{1.3cm}| p{2.1cm} | p{2.1cm} | p{1.6cm} |}
\hline
\textbf{State} & \textbf{$\mathbb{K}_{circ}, d_C=2$} & \textbf{$\mathbb{K}_{circ}, d_C=3$} & \textbf{$\mathbb{K}_{hw}$} \\
\hline
0000 &  0.8550 &  0.8379 &  0.9360 \\
\hline
0001 &  0.7988 &  0.7935 &  0.8774 \\
\hline
0010 &  0.7932 &  0.7773 &  0.8647 \\
\hline
0011 &  0.7378 &  0.7422 &  0.8174 \\
\hline
0100 &  0.7415 &  0.7136 &  0.8108 \\
\hline
0101 &  0.7046 &  0.6943 &  0.7739 \\
\hline
0110 &  0.6946 &  0.6846 &  0.7542 \\
\hline
0111 &  0.6450 &  0.6504 &  0.7209 \\
\hline
1000 &  0.8101 &  0.8115 &  0.8975 \\
\hline
1001 &  0.7783 &  0.7764 &  0.8486 \\
\hline
1010 &  0.7561 &  0.7500 &  0.8215 \\
\hline
1011 &  0.7332 &  0.7251 &  0.7812 \\
\hline
1100 &  0.7129 &  0.6858 &  0.7644 \\
\hline
1101 &  0.6802 &  0.6812 &  0.7419 \\
\hline
1110 &  0.7163 &  0.6667 &  0.7585 \\
\hline
\end{tabular}
\end{table}

When deployed on noisy hardware we can again construct an AEM that quantifies the probability of the target state being prepared, and the probabilities of other states being prepared.  In \Cref{sec:discussion} we calculated the Frobenius norm of the difference matrix $(\mathbb{1} - \mathbb{K})$ and showed that $\mathbb{K}_{circ}$ matrices are much farther from the identity matrix than $\mathbb{K}_{hw}$ matrices. The final AEM is assumed to be invertible and non-negative, but we note that the matrix is not symmetric.  Additionally the individual state fidelities are not uniform.  For $3$ AEM generated on the same day on the same set of qubits we report the individual basis state fidelities in \Cref{tab:state_fidelities}.